\documentclass[%
 aip,
 amsmath,amssymb,
 reprint,%
]{revtex4-1}

\usepackage{graphicx}
\usepackage{dcolumn}
\usepackage{bm}
\usepackage{ragged2e}   
\usepackage[normalem]{ulem}
\usepackage[most]{tcolorbox}

\usepackage[utf8]{inputenc}
\usepackage[T1]{fontenc}
\usepackage{mathptmx}
\usepackage{etoolbox}
\usepackage{subcaption}
\usepackage[english]{babel}
\usepackage{hyperref}

\makeatletter
\def\@email#1#2{%
 \endgroup
 \patchcmd{\titleblock@produce}
  {\frontmatter@RRAPformat}
  {\frontmatter@RRAPformat{\produce@RRAP{*#1\href{mailto:#2}{#2}}}\frontmatter@RRAPformat}
  {}{}
}%
\makeatother
\begin{document}

\preprint{AIP/123-QED}

\title{Fractals in rate-induced tipping}
\author{Jason Qianchuan Wang}
\thanks{These authors contributed equally to this work.}
 \affiliation{School of Mathematics and Statistics, The University of Sydney, Sydney, Australia}
\author{Yi Zheng}%
\thanks{These authors contributed equally to this work.}
\affiliation{School of Mathematics and Statistics, The University of Sydney, Sydney, Australia}
\author{Eduardo G. Altmann}%
\affiliation{School of Mathematics and Statistics, The University of Sydney, Sydney, Australia}
 \email{eduardo.altmann@sydney.edu.au}

\date{\today}

\begin{abstract}
When parameters of a dynamical system change sufficiently fast, critical transitions can take place even in the absence of bifurcations. This phenomenon is known as rate-induced tipping and has been reported in a variety of systems, from simple ordinary differential equations and maps to mathematical models in climate sciences and ecology. In most examples, the transition happens at a critical rate of parameter change, a rate-induced tipping point, and is associated with a simple unstable orbit (edge state). In this work, we explore how this simple picture changes when non-attracting fractal sets exist in the autonomous system, a ubiquitous situation in non-linear dynamics. We show that these fractals in phase-space induce fractals in parameter space, which control the rates and parameter changes that result in tipping. We explain how such rate-induced fractals appear and how the fractal dimensions of the different sets are related to each other. We illustrate our general theory in three paradigmatic systems: a piecewise linear one-dimensional map, the two-dimensional H\'enon map, and a forced pendulum. 
\end{abstract}

\maketitle

\begin{quotation}
Fractals appear naturally in different non-linear dynamical systems and provide a path to understand complex dynamics. In particular, the fractal dimension of invariant sets is one of the main properties characterizing chaotic dynamics. The question we consider here is whether fractals appear and play a similar role in the understanding of rate-induced tipping points, critical transitions that take place in non-autonomous dynamical systems. We present conditions under which fractals in phase-space induce fractals in parameter- and rate-space, thus controlling the boundary between tipping and non-tipping trajectories.
\end{quotation}

\section{Introduction}

{\it Tipping points} in climate and ecological models have renewed the interest in understanding dynamical processes leading to sudden changes in otherwise stable states~\cite{ashwin_tipping_2012,Okeeffe2020,Lohmann2021,Feudel2023}. Rate-induced tipping (R-tipping)~\cite{ashwin_tipping_2012} is a mechanism leading to such change in non-autonomous dynamical systems with time-dependent parameters, as observed and studied in a variety of systems in the last two decades~\cite{ashwin_tipping_2012,alkhayuon_rate-induced_2018,kaszas_tipping_2019,kiers_rate-induced_2020,Okeeffe2020,Lohmann2021,Feudel2023,wieczorek_rate-induced_2023,ritchie_rate-induced_2023,Lohmann2024,janosi_overview_2024,lohmann_risk_2021, ashwin_physical_2021, lohmann_predictability_2024, alkhayuon_weak_2020,panahi_global_2025}.
In R-tipping, the dynamics {\it tracks} the same stable state if parameters change slowly (small rate $r$), but {\it tips} out of it if the change is fast (large rate $r$). The tipping point is associated with a critical rate ($r_c$). In the simplest and best studied examples of R-tipping, this critical rate is at the tracking-tipping boundary and leads to trajectories approaching an {\it edge state} that is an unstable fixed point~\cite{ashwin_tipping_2012,kiers_rate-induced_2020} or periodic orbit~\cite{wieczorek_rate-induced_2023,alkhayuon_rate-induced_2018} of the autonomous dynamics.

R-tipping is an example of the important role played by unstable orbits in non-linear systems, even if these orbits are not directly observed. Similarly, in the study of {\it transient chaos}~\cite{lai_transient_2011,aguirre_fractal_2009,altmann_leaking_2013}, fractal invariant saddles are the central object of interest, as they control the most relevant dynamical and observable quantities: the escape rate of the saddle controls the rate at which trajectories escape or transition between attractors; the stable manifold of the saddle often acts as a fractal basin boundary between attractors or escape directions; and the fractal dimension of the saddle can be connected to both the escape rate and the Lyapunov exponent of the transient chaotic dynamics. Importantly, such fractal sets appear in many different non-linear dynamical systems without the need for fine-tuning of parameters, as illustrated by the wide range of examples reviewed in the literature~\cite{lai_transient_2011,aguirre_fractal_2009,altmann_leaking_2013}.

Previous studies have investigated several ways in which fractal invariant sets in the phase-space of dynamical systems affect R-tipping. These include R-tipping involving chaotic or fractal attractors~\cite{Kaszas2016,ashwin_physical_2021,lohmann_predictability_2024, alkhayuon_weak_2020}, and the role of fractal saddles and basin boundaries in producing sensitive dependence of tipping outcomes, affecting tipping probabilities, and limiting predictability~\cite{kaszas_tipping_2019, lohmann_predictability_2024, lohmann_risk_2021}. In particular, chaotic or fractal edge states have been reported in different climate models~\cite{Lucarini2017,Lohmann2021,Lohmann2024, lohmann_risk_2021, ashwin_physical_2021, lohmann_predictability_2024, mehling_limits_2024}. Recently, the importance of further investigating the effect of fractal saddles on R-tipping has been emphasized~\cite{wieczorek_rate-induced_2023,Feudel2023}.

In this paper, we show how transient chaos and fractal saddles create an extreme sensitivity of the asymptotic state (track or tip) on the choice of parameter change and its rate $r$. We propose a generic mechanism connecting fractal basin boundaries in phase-space to fractal R-tipping in parameter and rate spaces. We first demonstrate this behavior through simulations of simple dynamical systems. We then develop a heuristic theory and identify general conditions under which the fractal co-dimensions of these three fractals are expected to be equal.

The paper is divided as follows. We start in Sec.~\ref{sec.formalism} with the definitions and mathematical formalism for the study of R-tipping in (discrete-time) dynamical systems~\cite{kiers_rate-induced_2020}. In Sec.~\ref{sec.examples}, we investigate three systems showing fractal R-tipping, including numerical evidence of the equality of the fractal co-dimension of the three different sets. This motivates the introduction, in Sec.~\ref{sec.theory},  of a general mechanism that explains the numerical observations. Finally, we conclude and discuss our results in Sec.~\ref{sec.conclusion}. Detailed derivations and numerical simulations of additional cases are reported in the Appendices and the code to re-produce our results is available in our repository~\cite{wang_fractal_tipping_2026}.

\section{General formalism}\label{sec.formalism}

Here we investigate generic R-tipping phenomena following the discrete-time formalism introduced in Ref.~\cite{kiers_rate-induced_2020}. The autonomous dynamics is defined by a $\mathbb{C}^0$ (continuous) function $F$ with parameters $\lambda \in \mathbb{R}^m$ that maps points $x\in \Omega$ from and to the phase-space $\Omega \subseteq \mathbb{R}^d$ as
\begin{equation}\label{eq.f}
x_{n+1} = F(x_n; \lambda),
\end{equation}
where $n \in \mathbb{Z}$ is the discrete time. The non-autonomous dynamics is obtained by changing the parameters $\lambda$ in time as
\begin{equation}\label{eq.map}
\begin{aligned}
x_{n+1} &= F(x_n; \Lambda(s_n)),\\
s_n &= r\,n .
\end{aligned}
\end{equation}
where $r\ge0$ is the {\it rate} of change of $\lambda$ and $\Lambda(s)=\Lambda(s_n)$ is the protocol specifying the path of parameter change, which we consider a $\mathbb{C}^0$ (continuous) function $\mathbb{R}\mapsto \mathbb{R}^m$. We will consider different paths $\Lambda(s)$ such that $\lambda_0=\Lambda(s=0)$ (at $n=0$) and
\begin{equation}\label{eq:lambda_asymptote}
\lambda_{\pm} = \lim_{s\rightarrow \pm\infty} \Lambda(s).    
\end{equation}

We consider autonomous dynamical systems $F$ that show multi-stability, i.e., for which different initial conditions in Eq.~(\ref{eq.f}) lead to more than one asymptotic ($n\rightarrow \infty$) state (e.g., attractors, unbounded solutions). For simplicity, in the examples we consider below one of such asymptotic states is a stable fixed point $x^*(\lambda)$ and we study whether trajectories starting at $x^*(\lambda)$ track or tip from $x^*(\lambda)$ as the parameter $\lambda$ changes according to $\Lambda(s)$. As we are interested exclusively in R-tipping, rather than in tipping due to bifurcations, we focus on parameter changes that follow a {\it stable paths}, i.e., parameter paths $\Lambda(s)$ so that $x^*(\lambda)$ is stable for all $\lambda \in \Lambda(s)$. 
As to initial conditions $x_0$, we primarily consider trajectories starting at the fixed point $x^*(\Lambda(s))$, either at $s=n=0$ or at $s\rightarrow -\infty$, depending on the choice of $\Lambda(s)$.  If asymptotically the trajectory approaches the stable fixed point $x^*(\lambda_+)$ of the frozen system $F(x, \lambda_+)$, we say that it {\it tracks}. Otherwise,  we say that it {\it tips} (e.g., if the trajectory approaches another attractor of $F(x,\lambda_+)$ or diverges). We thus define the tipping function $\phi$ as

\begin{equation}\label{eq.phi}
\phi(r,x_0;\Lambda(s)) = 
    \begin{cases}
    \text{ track} &\text{ if } \lim\limits_{n\rightarrow \infty} x(n) = x^*(\lambda_+),\\
    \text{ tip} &\text{ otherwise} ,
    \end{cases}
\end{equation}
where $x(n)$ is obtained by iterating Eq.~\eqref{eq.map} with initial condition $x^*(\Lambda(s))$ at $s=0$ or $s\rightarrow -\infty$ depending on $\Lambda(s)$. Contrary to alternative definitions~\cite{wieczorek_rate-induced_2023,Feudel2023}, $\phi$ in Eq.~\eqref{eq.phi} considers only the $n\rightarrow \infty$ behavior of $x$ and classifies as tracking the trajectories that depart the vicinity of $x^*(\lambda(n))$ but that eventually settle into it. 
We are primarily interested in understanding how variations of $r$, $x_0$, and $\Lambda(s)$ lead to a changes in the outcome of $\phi$. The parameter values at which this happens will be denoted critical or boundary values. 

Since the parameters $\lambda$ are changing in a stable path $\Lambda(s)$ and the initial condition is in $x^*$, for sufficiently small rate $r$ the trajectory tracks~\cite{kiers_rate-induced_2020}. 
As the rate of parameter change $r$ grows, the trajectory {\it may} tip. A rate $r=r_c$ is called critical if it is at the boundary between tracking and tipping, i.e., if every open interval around $r_c$ contains both $\phi = \text{track}$ and $\phi = \text{tip}$. In particular, we denote the smallest critical $r$ as $r_{c1}$. In the traditional (simplest) examples of R-tipping, there is only one such point $r_c$ (tipping point), and it is associated with an unstable fixed point -- an {\it edge state} -- that is approached when $r=r_c$.  In order to observe fractals in R-tipping, the key idea we follow~\cite{Lohmann2021,wieczorek_rate-induced_2023,Feudel2023}  is to explore the consequence of fractal saddles acting as edge states to the transition between tracking and tipping, as quantified by the dependence of the tipping function~$\phi(r,x_0;\Lambda(s))$ -- defined in Eq.~(\ref{eq.phi}) -- on the tipping related parameter $r$ and the path $\Lambda(s)$.

\section{Examples}\label{sec.examples}

We now investigate the effect of fractal saddles in R-tipping using three simple maps $F_A,F_B,$ and $F_C$ as $F$ in Eq.~(\ref{eq.map}).

\subsection{Piecewise-linear map ($d=1$)}

We start by constructing a simple example of dynamical system showing fractal R-tipping. Our starting point is the open tent map
\begin{eqnarray}\label{eq:tent}
    x_{n+1} &= & 
    \begin{cases}
        \mu \; x_n, &x_n < 1/2, \\
        \mu \; (1-x_n), & x_n \ge 1/2,
    \end{cases}
\end{eqnarray}
where $x\in \mathbb{R}$ and $\mu \ge 2$ is a parameter. For $\mu=2,$ the closed tent map is recovered, which leads to a fully chaotic dynamics in $x\in[0,1]$. For $\mu>2$, with probability one, randomly selected initial conditions $x_0 \in [0,1]$ diverge to $-\infty$ as $n \to \infty$. Nevertheless, there is a Lebesgue measure zero set of non-escaping points that build a fractal repeller. For $\mu=3$, which we use throughout this paper, the repeller is the middle-third Cantor set \cite{aguirre_fractal_2009}. 

\begin{figure}[!h]
        \begin{center}
        \includegraphics[width=0.8\linewidth]{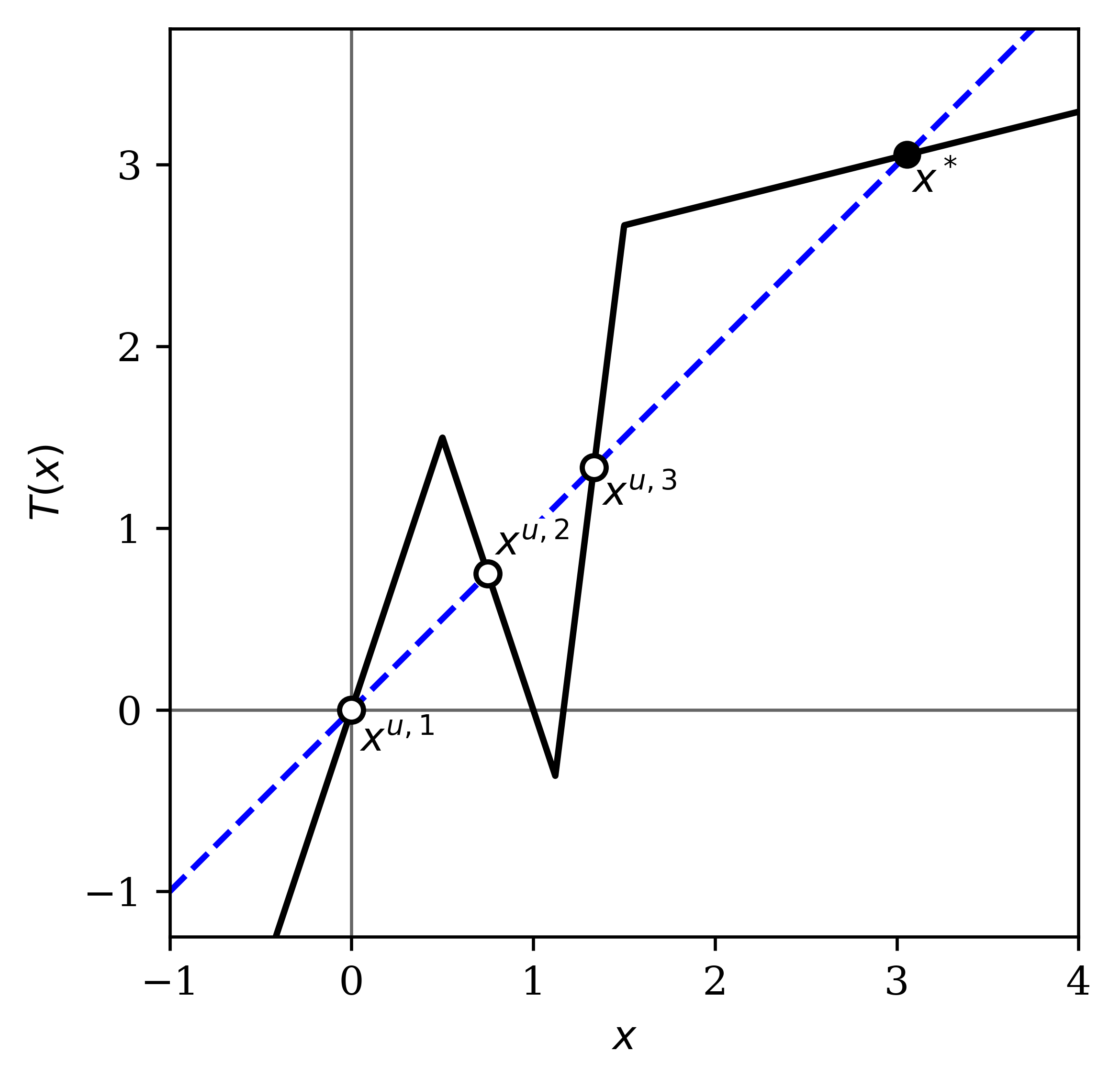} 
        \end{center}
    \caption{\justifying \justifying Graphical representation of the modified tent map $T(x)$ in Eq.~(\ref{eq.example1}). 
    There are four fixed points at the intersection of $T(x)$ (solid line) and the diagonal $y=x$ (dashed blue line): two unstable as in the open tent map ($x^{u,1}=0$ and $x^{u,2}=3/4$), a third unstable ($x^{u,3}=4/3$) in the intermediate interval ($37/33 < x < 3/2$), and one stable ($x^*=55/18$). We investigate R-tipping out of $x^*$ by shifting this map along the diagonal -- see Eqs.~\eqref{eq.lambdaTent}-\eqref{eq:tanh_drift}.}
    \label{fig:modified_tent}
\end{figure}

\begin{figure}[!h]
    \centering

\begin{subfigure}[t]{0.46\textwidth}
    \centering
    \caption{\justifying}
    \vspace{0pt}

    \begin{minipage}[t][0.98\linewidth][t]{\linewidth}
        \centering
        \includegraphics[width=0.93\linewidth]{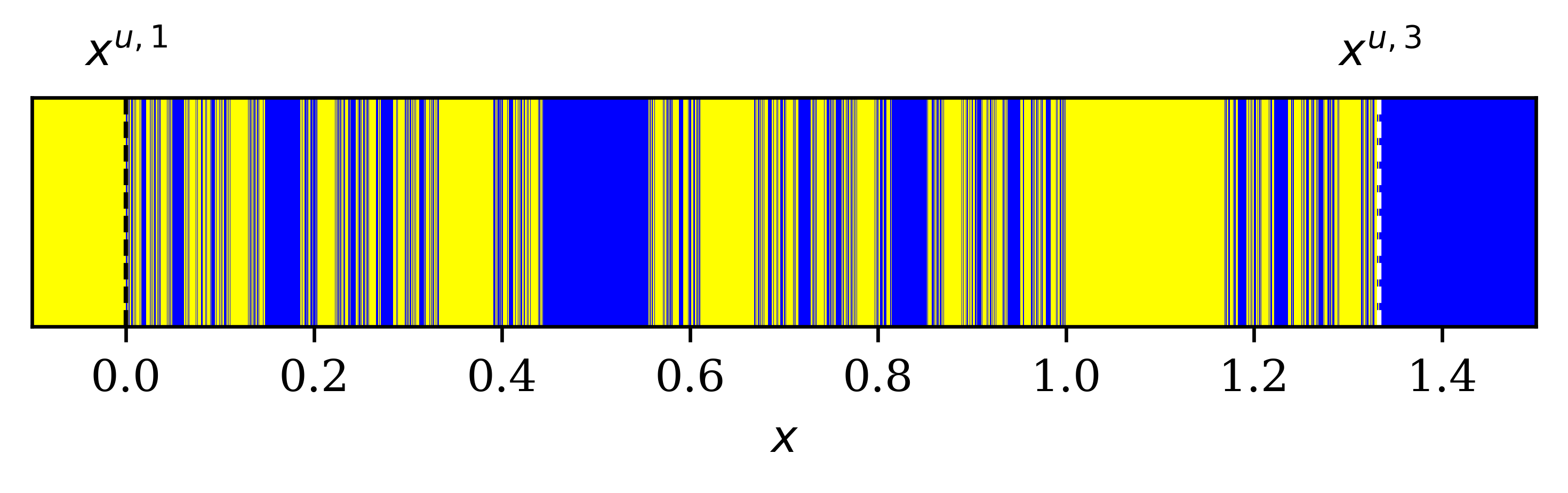}\vfill
        \includegraphics[width=0.98\linewidth]{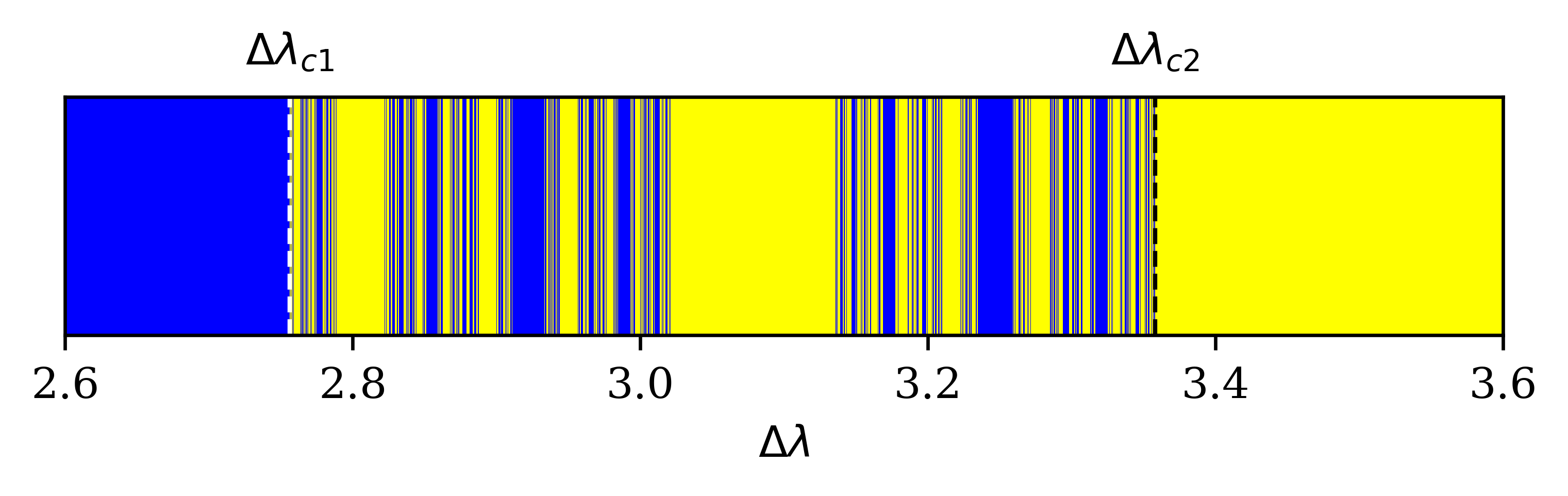}\vfill
        \includegraphics[width=\linewidth]{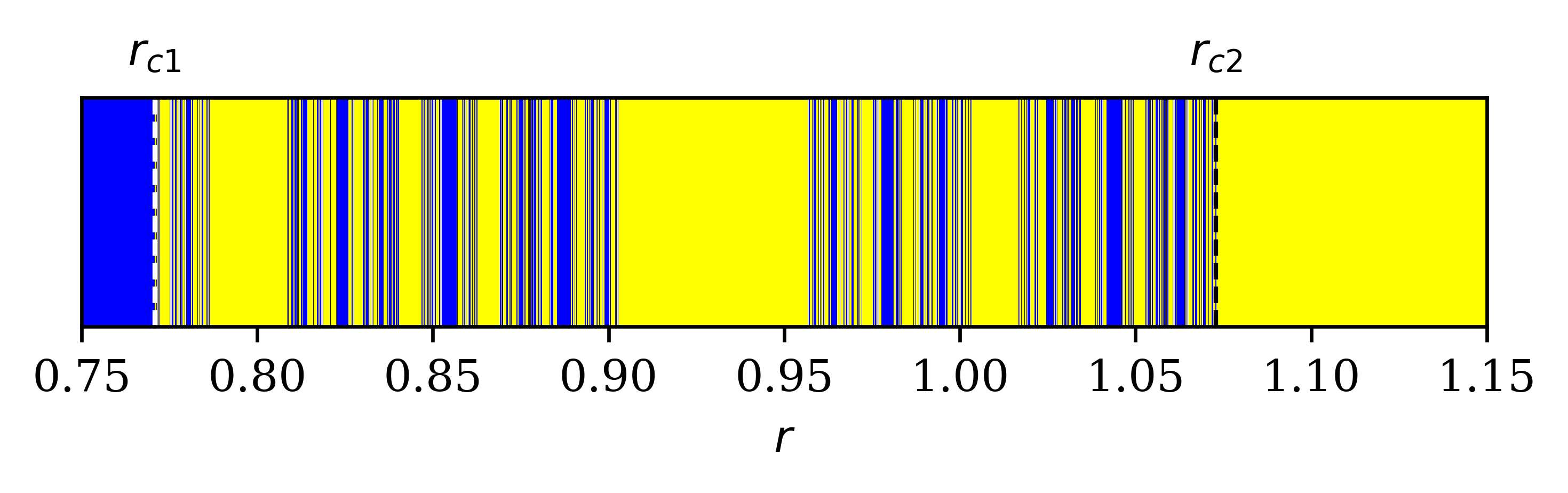}
    \end{minipage}

    \label{fig:fractals}
\end{subfigure}
    \hfill
    \begin{subfigure}{0.46\textwidth}
        \centering
        \caption{\justifying }
        \includegraphics[width=1\linewidth]{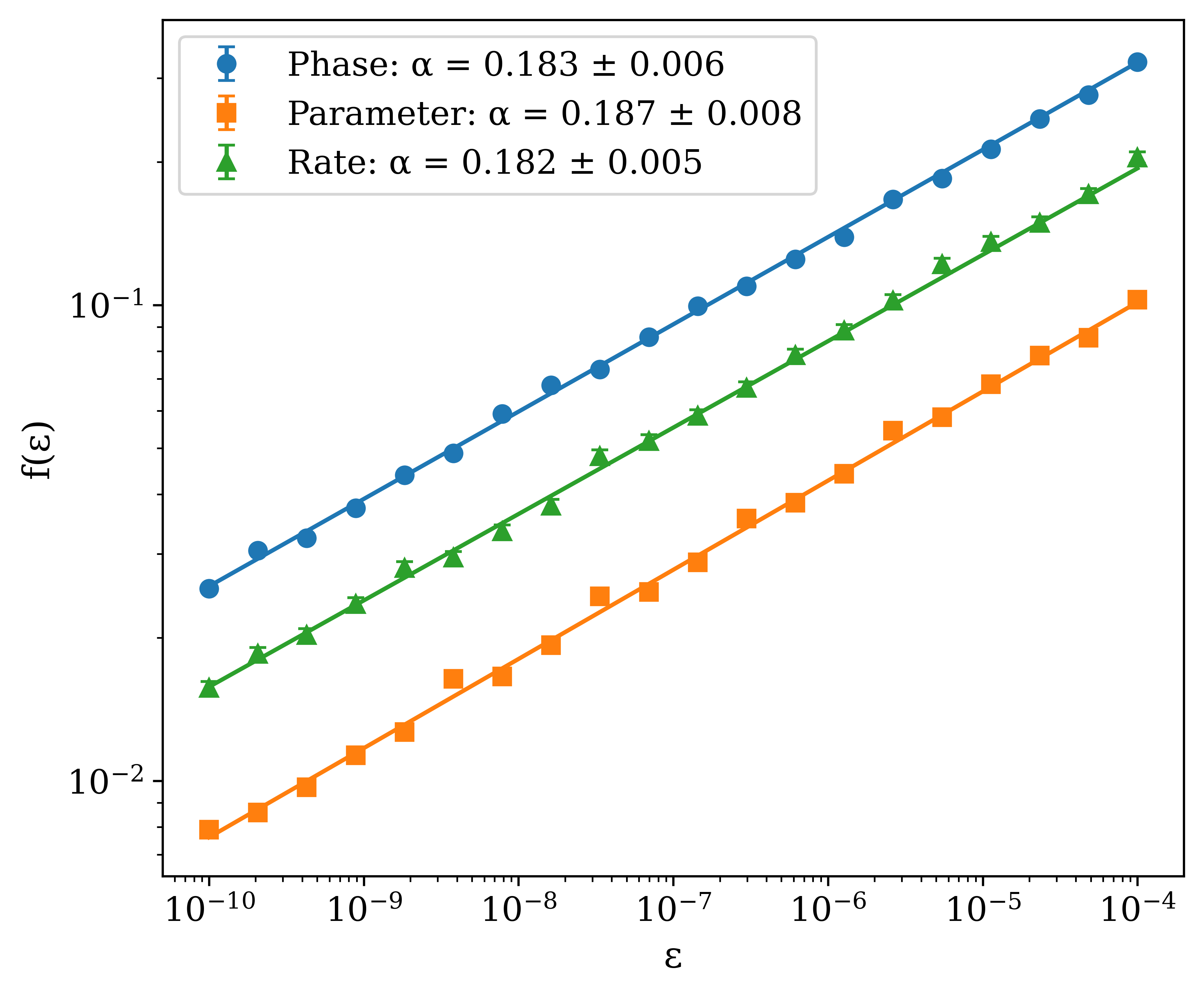}
        \label{fig:fits}
    \end{subfigure}
    \caption{\justifying The three fractals in the piecewise-linear map~(\ref{eq.lambdaTent}) have the same co-dimension $\alpha$. (a) Plot of the function $\phi(r,x_0;\Lambda(s))$ in Eq.~(\ref{eq.phi}) -- with blue corresponding to $\phi=$track (the fixed point $x^*$) and yellow to $\phi=$tip $(x\rightarrow-\infty$) --  as a function of different parameters (top to bottom): initial conditions $x=x_0$ (with $r=\lambda_+=0$, the phase-space fractal); range of parameter variation $\Delta \lambda = 2\lambda_+$ used in $\Lambda(s)$ (with $r\rightarrow \infty$, the parameter-space fractal); and rate of change $r$ (with $\Delta \lambda = 4$, the rate-space fractal).  (b) Estimation of the fractal co-dimension $\alpha$ of the track/tip boundary for the three fractals shown in panel (a), see Appendix~\ref{app.fractal} for details on the computation. Points for the phase-space fractal are multiplied by 2 for clarity.}
    \label{fig:combined}
\end{figure}

We now modify the tent map in Eq.~\eqref{eq:tent} to obtain a new map $T(x)$ in which R-tipping can take place. To introduce multi-stability, we require $T$ to have a fixed point $x^*=T(x^*)$ that is stable ($|T'(x^*)|<1$) and on the right of the domain of the tent map, i.e., $x^*>1$. A simple modification of Eq.~\eqref{eq:tent} that achieves this, retaining $T(x)$ as a piecewise linear continuous function and using $\mu=3$, is to add two new branches as
\begin{eqnarray}\label{eq.example1}
    x_{n+1} = T(x_n) &=&
\begin{cases}
        3x_n, &x_n < 1/2 \\
        3(1-x_n), &  1/2 \le x_n < \frac{37}{33} \\
8 x_n-\frac{28}{3}, & \frac{37}{33} \leq x_n < \frac{3}{2}, \\
\frac{1}{4}x_n+\frac{55}{24}, & x_n \geq \frac{3}{2}.
\end{cases}
\end{eqnarray}
This function is plotted in Fig.~\ref{fig:modified_tent}. Initial conditions in the last branch ($x \ge 3/2$) converge to the stable fixed point $(x^*=55/18=3.0\bar{5})$ due to the choice of its slope $T'(x^*)=1/4<1$. The remaining parameters of the two branches were chosen in such a way that the interval $x\in[1,3/2]$ is split in three equal sized sub-intervals such that $x\in[1,7/6]$ leads immediately to escape (i.e., $T(x) < 0 \Rightarrow x\rightarrow -\infty$), $x\in[7/6, 4/3]$ leads to a re-injection in the tent map domain (i.e., $T(x) \in [0,1]$), and $x\in[4/3, 3/2]$ leads to the stable fixed point (i.e., $x\rightarrow x^*)$. 

We now consider non-autonomous changes of the map~(\ref{eq.example1}) that can induce R-tipping out of the stable fixed point $x^*$. As in Ref.~\cite{kiers_rate-induced_2020} (Example 4.2), we introduce a new map $F_A$ with a parameter $\lambda$ that preserves the dynamics and simply shifts $T(x)$ along the line $y=x$ as
\begin{equation}\label{eq.lambdaTent}
x_{n+1}=F_A(x_n;\lambda) = T(x_n-\lambda)+\lambda,
\end{equation}
with $T$ given by Eq.~(\ref{eq.example1}), see Appendix~\ref{sec.appnedixMap1} for an explicit expression of $F_A$ for arbitrary parameters. In particular, the four fixed points of $T(x)$ appear as fixed points of $F_A(x)$ shifted by $\lambda$, i.e., $x^*+\lambda$ is a stable and $x^{u,i}+\lambda$ ($i=1,2,3$) are unstable fixed points of $F_A(x,\lambda)$. 

The map $F_A$ is a $d=1$ example of the map $F$ in Eq.~(\ref{eq.map}). The non-autonomous system~(\ref{eq.map}) is obtained by specifying a protocol $\Lambda(s)$ for the parameter change, which here we use 
\begin{equation}\label{eq:tanh_drift}
    \Lambda_A(s) = \lambda_+\tanh(s) = \lambda_+ \tanh(rn),
\end{equation}
and initial condition $x=x^*(\lambda_-=-\lambda_+)$ at $n\rightarrow -\infty$ so that the trajectory $x_n$ corresponds to the pullback attractor of $x^*$ as in Ref.~\cite{kiers_rate-induced_2020}.

\begin{figure*}[!t]
    \centering

    \begin{subfigure}{0.3\textwidth}
        \centering
        \caption{\justifying }
        \includegraphics[width=\linewidth]{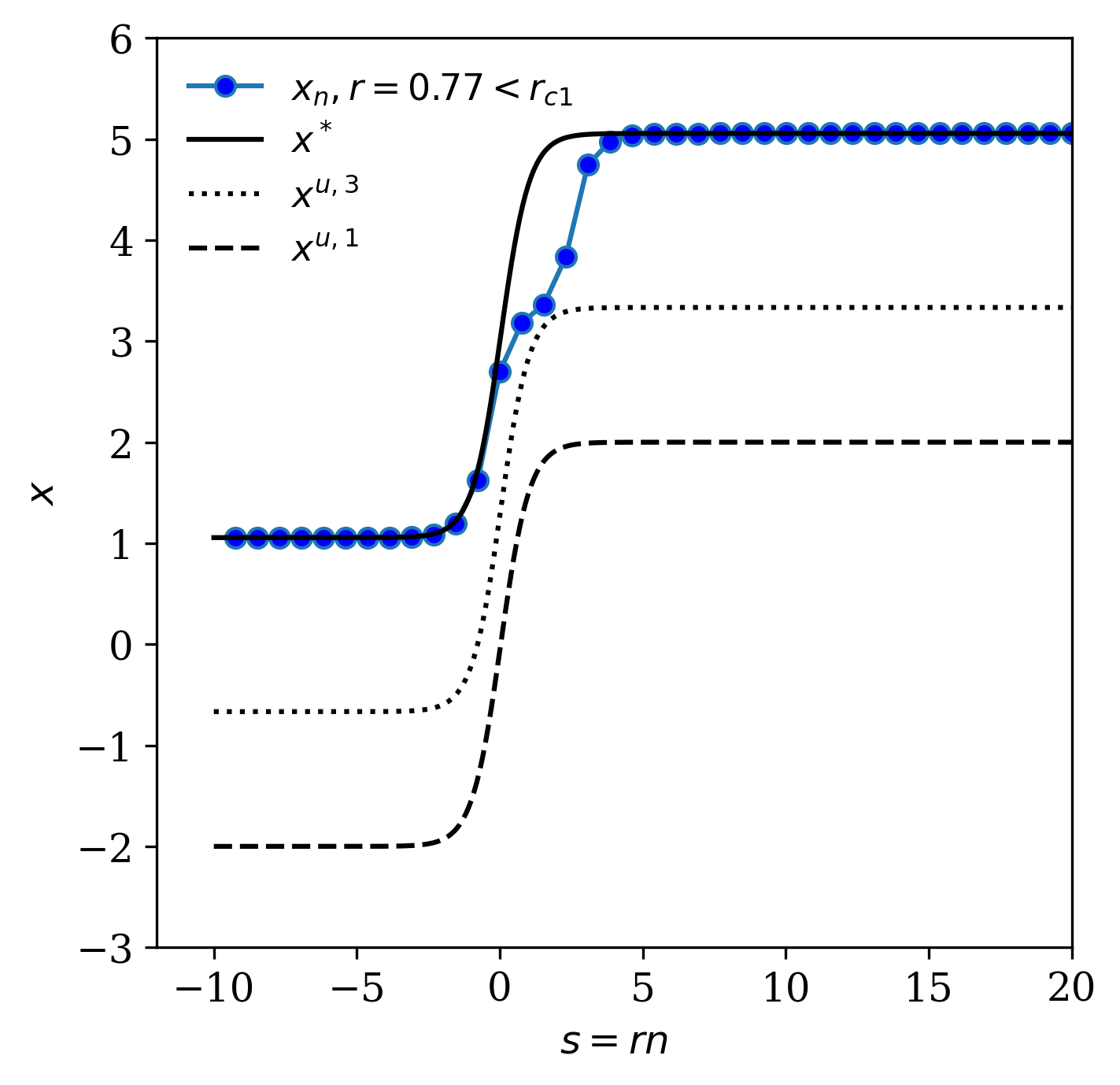}
        \label{fig:rate_tracking}
    \end{subfigure}
    \hfill
    \begin{subfigure}{0.3\textwidth}
        \centering
        \caption{\justifying }
        \includegraphics[width=\linewidth]{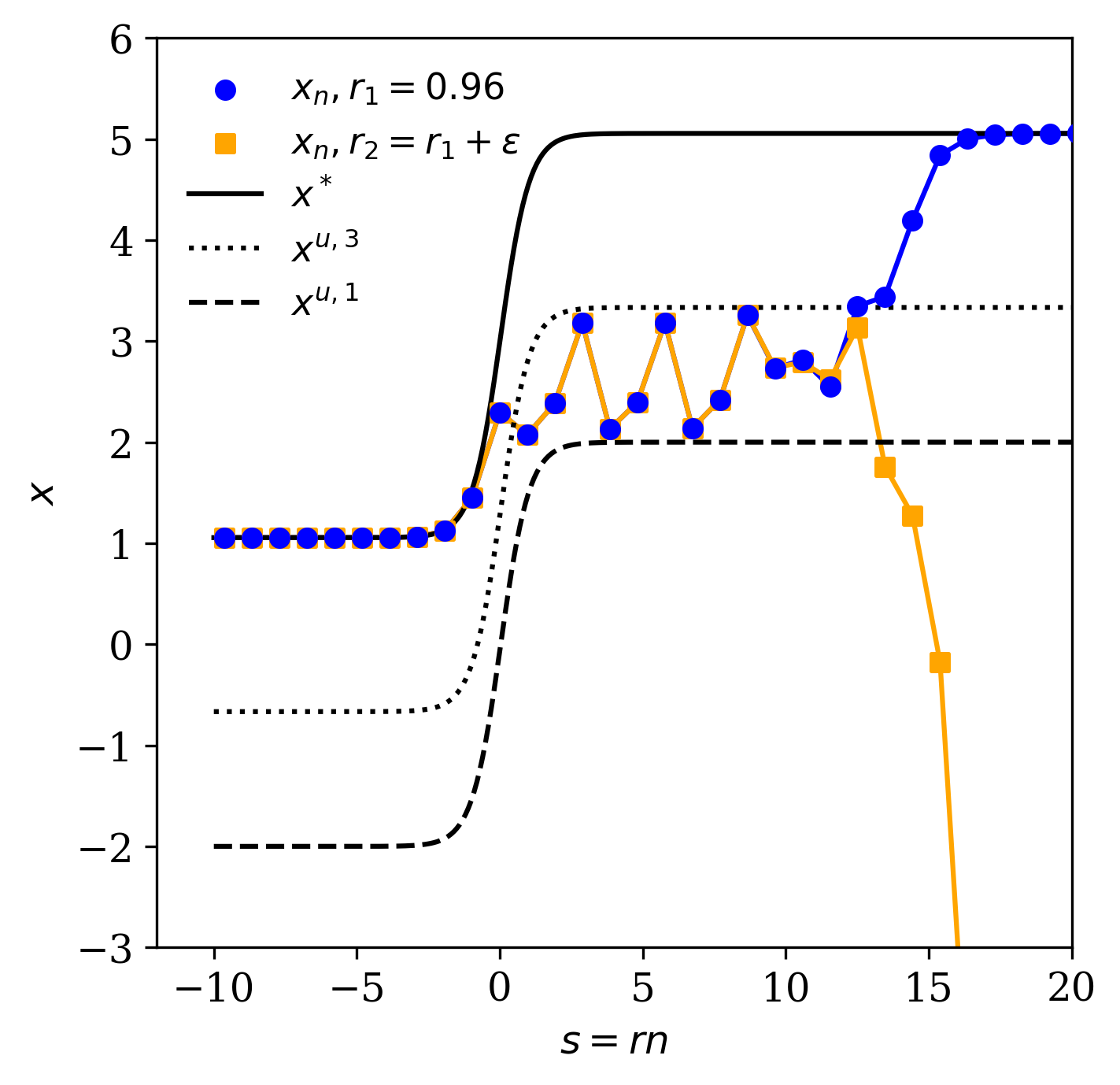}
        \label{fig:rate_imtermediate}
    \end{subfigure}
    \hfill
    \begin{subfigure}{0.3\textwidth}
        \centering
        \caption{\justifying }
        \includegraphics[width=\linewidth]{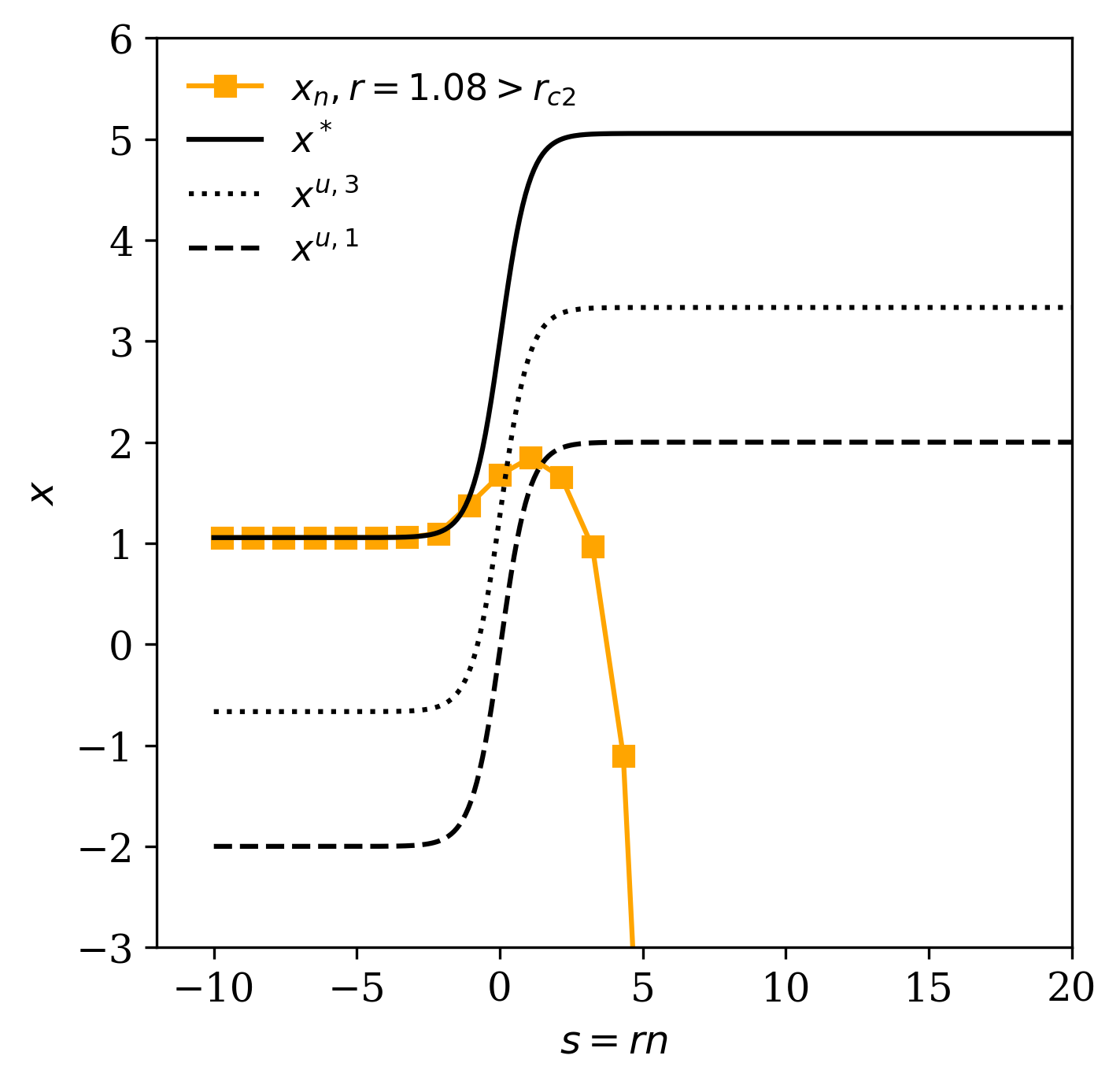}
        \label{fig:rate_tipping}
    \end{subfigure}
    
    \caption{\justifying Tracking and tipping trajectories in the non-autonomous piecewise-linear map defined by Eqs.~(\ref{eq.example1})-\eqref{eq:tanh_drift}. The three panels show the frozen paths of the stable -- $(x^*(\Lambda(s)),s)$ as solid line -- and unstable fixed points -- $(x^{u,1}(\Lambda(s)),s)$ and $(x^{u,3}(\Lambda(s)),s)$  as dashed and dotted black lines, respectively --, and the trajectory $x_n$ obtained starting at $x^*$ for $n\rightarrow -\infty$ and evolving using Eq.~(\ref{eq.map}) with $F=F_1$ -- in Eq.~(\ref{eq.lambdaTent}) -- and $\Lambda(s)$ -- in Eq.~(\ref{eq:tanh_drift}) -- with parameters $\lambda_+=2$ and different values of the rate $r$ of parameter change: (a) small rate $r= 0.77  < r_{c1} \approx 0.7708$, leading to a tracking trajectory; (b) two similar rates $r_1=0.96243528753$ and $r_2=r_1+10^{-9}$ close to a critical rate $r_{c1}<r_c<r_{c2}$ (boundary point), leading to both tracking and a tipping trajectories; and (c)  large rate $r=1.08 > r_{c2} \approx 1.0727$, leading to a tipping trajectory.}
    \label{fig:rates}
\end{figure*}

We now investigate R-tipping in the dynamical system~\eqref{eq.map} defined by Eqs.~\eqref{eq.example1}-\eqref{eq:tanh_drift}. We focus on the dependence of the tipping function $\phi$ in Eq.~\eqref{eq.phi} on $\Delta \lambda = \lambda_+-\lambda_- = 2\lambda_+$ and $r$ in Eq.~(\ref{eq:tanh_drift}), which together uniquely specify the parameter change.  For sufficiently small $r<r_{c1}$ and $\Delta \lambda <\Delta \lambda_{c1}$, $\phi=$ track because the parameter variation is in a stable path. 
  The small $r,\Delta \lambda$ regime is valid up to the first critical values $\Delta \lambda_{c1}\approx2.756$ and  $r_{c1} \approx 0.771$, which can be computed considering the first time trajectories $x_n$ cross the frozen path ($x^{u,3}(\Lambda(s)),s)$ of the largest unstable fixed point $x^{u,3}$. Similarly, for sufficient large $\Delta \lambda\ge\Delta \lambda_{c2}\approx3.358$ and $r\ge r_{c2} \approx 1.073$, the trajectory crosses the frozen path of the smallest unstable fixed point $(x^{u,1}(\Lambda(s)),s)$ and therefore $\phi=$ tip (the critical values reported above were estimated numerically). We are also particularly interested in the case $\Delta \lambda$ for $r\rightarrow \infty$ because the appearance of $\phi$= tip for any $\Delta \lambda$ in this case is a {\it sufficient} condition for the appearance of R-tipping (see Ref.~\cite{kiers_rate-induced_2020}, Theorem 4.9). Figure~\ref{fig:combined} shows numerical results of the dependence of $\phi$ on $x$, $\Delta \lambda$, and $r$ (see also Appendix~\ref{sec:two_params} for plots of $\phi$ as a function of $r$ and $\Delta \lambda$). They confirm the prediction of track (tip) for parameters  below (above) the first (second) critical value $r_{c1}$ ($r_{c2})$. More interestingly, the numerical results show that in between these values, there is an intricate variation between $\phi=$ track and $\phi=$ tip.  This observation 
was previously noted in Ref.~\cite{lohmann_risk_2021} in an ocean circulation model, which also displayed a fractal basin boundary in the phase-space and a non-monotonic dependence of the tipping outcome~$\phi$ on the rate $r$. In our one-dimensional map, which also has a fractal basin boundary, we observe that  $\phi$ changes with $r$ on a fractal set. Remarkably, this rate-space fractal has the same co-dimension as the fractals obtained by varying the initial condition $x$ (for $r=\Delta \lambda=0$, fractal repeller of the autonomous map) and $\Delta \lambda$ (for $r\rightarrow \infty)$. 
  Each fractal co-dimension $\alpha$ is computed as the slope of the scaling between the fraction of $\varepsilon$-uncertain points $f(\varepsilon)$ and $\varepsilon$, using the uncertainty algorithm~\cite{McDondald1985,lai_transient_2011} described in detail in Appendix~\ref{app.fractal}.
  
We now explain the connection between the three fractal observations reported above. The repeller of the autonomous map~\eqref{eq.example1} (the phase-space fractal) is composed of the union of the middle third Cantor set of the tent map~\eqref{eq:tent} and other components in $x\in[0,4/3]\equiv[x^{u,1},x^{u,3}]$ created by the third linear branch in Eq.~\eqref{eq.example1} due to re-injection. This fractal can be directly connected to the fractal in $\Delta \lambda$ (for $r\rightarrow \infty$, the parameter-space fractal) by noting that, due to the simplicity of the effect of $\lambda$ in the dynamics specified in Eq.~\eqref{eq.lambdaTent}, a variation $\Delta \lambda$ is equivalent to the autonomous map, Eq.~\eqref{eq.example1}, initialized at $x_0 = F_A(x^* - \lambda_+;0)$, i.e., at the one-time iteration (with parameter $\lambda=\tanh(0)=0$) of the fixed point $x^*$ of the map $F_A$ with parameter $\lambda_-=-\lambda_+$. This provides a direct connection of the parameter-space fractal to the phase-space fractal.
The connection to the fractal in $r$ (fixed $\Delta \lambda > \Delta \lambda_{c1}$, the rate-space fractal) to the other fractals is more subtle and will be fully discussed in the next section. An intuitive explanation is that for values $r_{c1} < r < r_{c2}$ the trajectories enter the region $x^{u,1} < x < x^{u,3}$ and experience a transiently chaotic dynamics. The closer $r$ is to any critical rate $r_c$, the longer $x_n$ spends between the (frozen path of the) unstable periodic orbits reproducing closely the dynamics of the fractal repeller (at $\lambda=\lambda_+$). This picture is confirmed in Fig.~\ref{fig:rates}, which shows the dynamics of tracking and tipping trajectories for different $r$'s.

\subsection{H\'enon map ($d=2$)}

\begin{figure}[!t]
    \begin{subfigure}{0.45\textwidth}
        \centering
        \caption{\justifying }
        \includegraphics[width=0.9\linewidth]{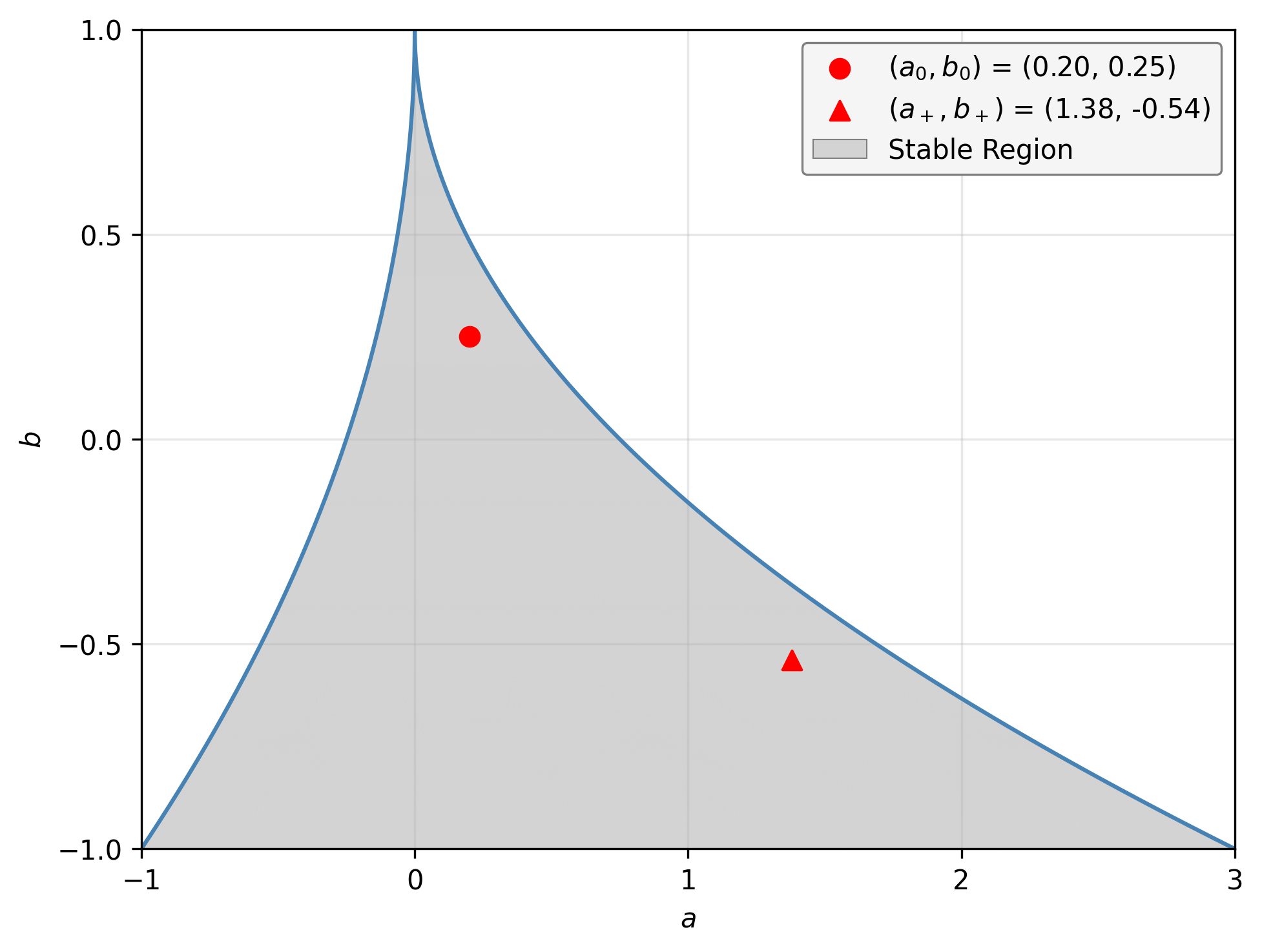} 
        \end{subfigure}
        
    \begin{subfigure}{0.45\textwidth}
        \centering
        \caption{\justifying }
            \includegraphics[width=0.4\linewidth]{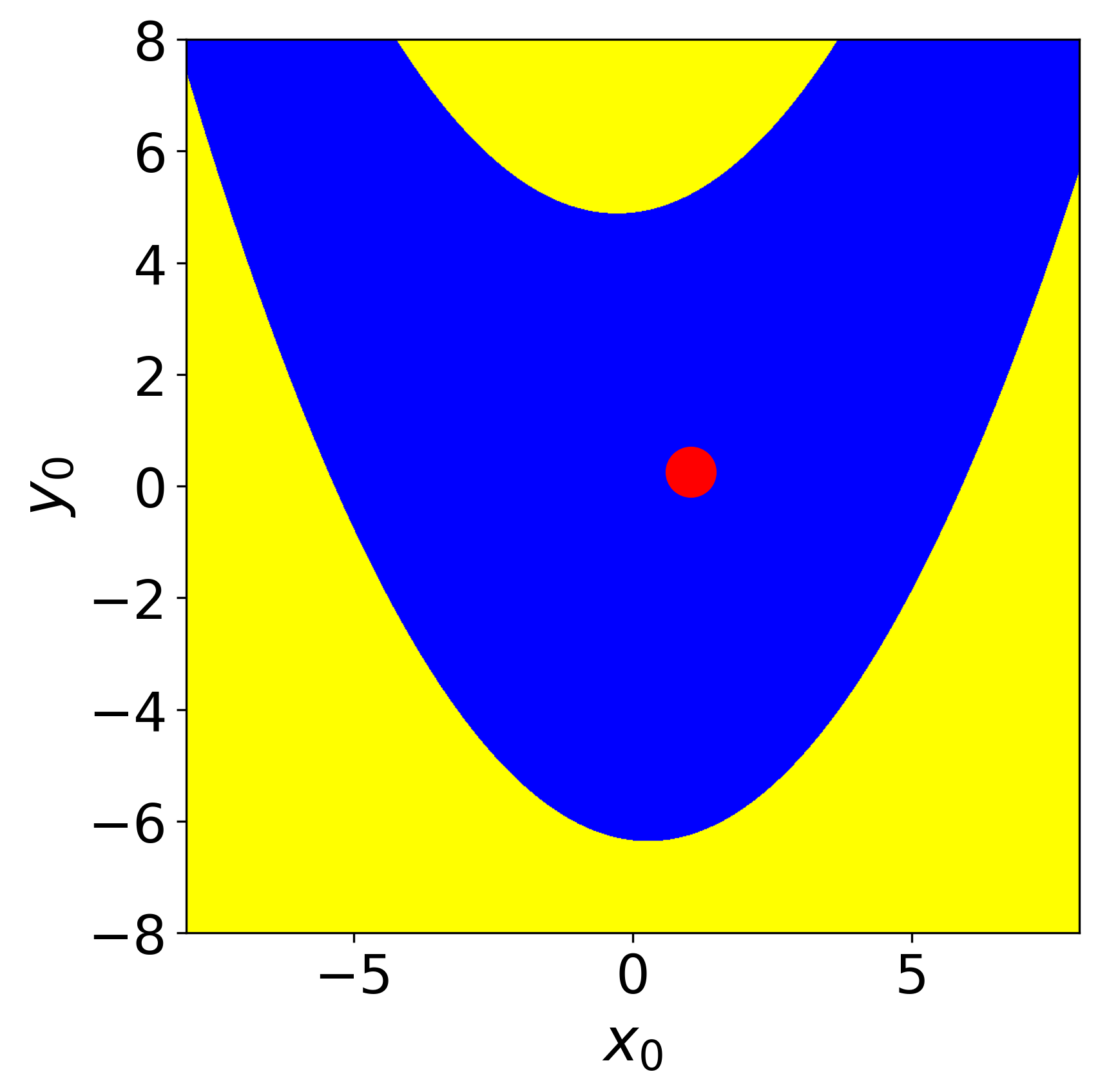}
            \includegraphics[width=0.41\linewidth]{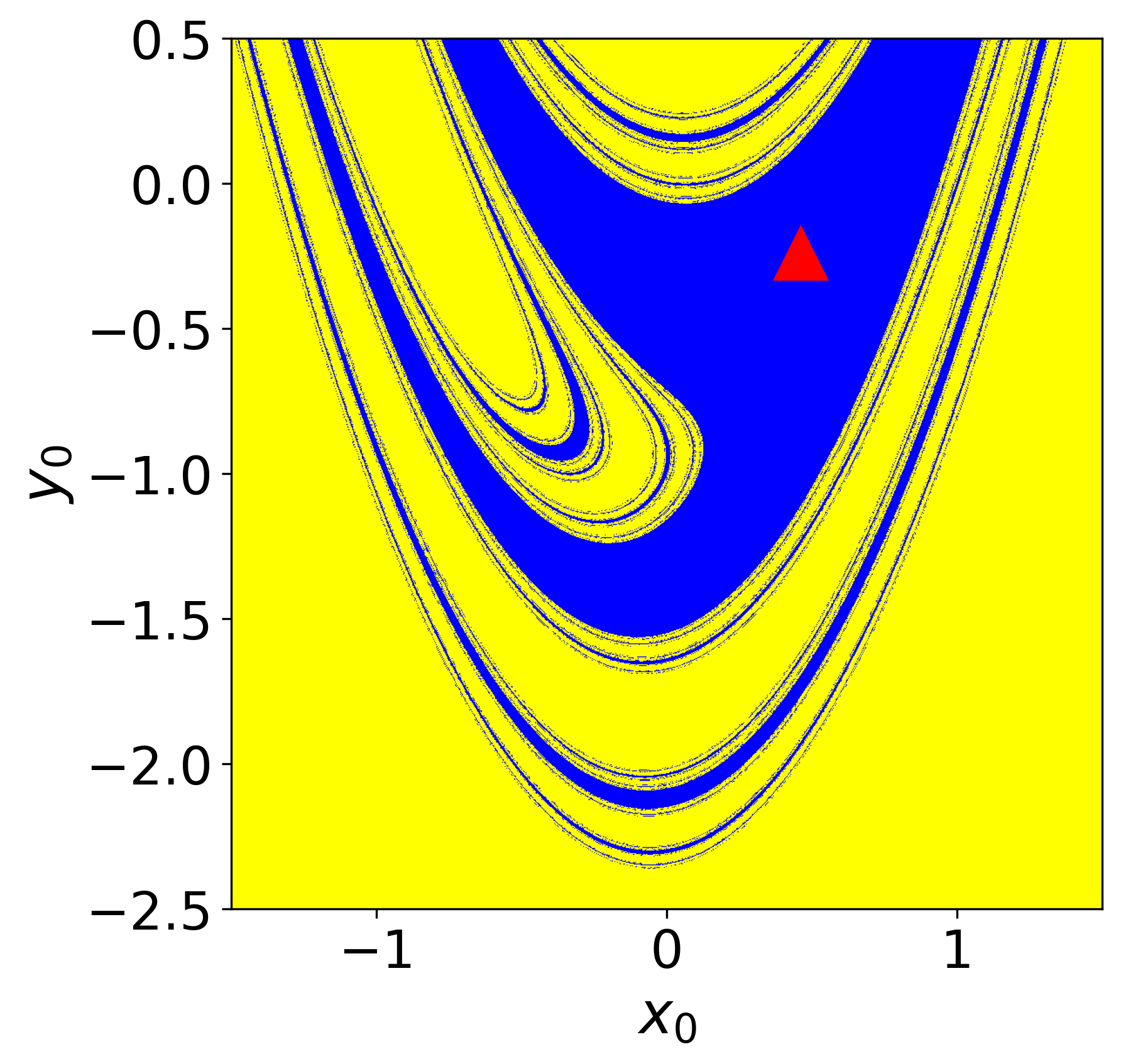}  
        \end{subfigure}
        \caption{\justifying The basin of attraction of the fixed point of the H\'enon map can have smooth or fractal boundaries. 
        (a) Parameter space $\lambda=(a,b)$ of the H\'enon map~\eqref{eq:henon}. The gray shaded region corresponds to the region $\Gamma$ for which the fixed point~(\ref{eq.henonfixedpoint}) is stable, defined in Eq.~(\ref{eq.Gamma}). The two highlighted points indicate 
        the values $\lambda_0=(a_0,b_0)  = (0.20, 0.25)$ and $\lambda_+=(a_+,b_+)=(1.38, -0.54)$ used in our investigation (see legend). (b) The basin of attraction (in blue) of the stable fixed point (red symbol) for $\lambda=\lambda_0$ (left, smooth 
        boundary) and $\lambda=\lambda_+$ (right, fractal boundary). Initial conditions in yellow diverge to infinity as $n\rightarrow +\infty$: $(x_n,y_n) \rightarrow (-\infty,-\infty)$ for $\lambda_0$; and  $(x_n,y_n) \rightarrow (-\infty,+\infty)$ for $\lambda_+$).}
        \label{fig:henonStableRegionAndBasin}\label{fig.henon1}
\end{figure}

Our second example is chosen to show that the appearance of fractals in R-tipping does not require a carefully designed dynamical system. We thus consider one of the most-studied simple systems with complex dynamics, the H\'enon map 
\begin{align}\label{eq:henon}
   F_B= \begin{cases}
        x_{n+1} &= 1-ax_n^2+y_n, \\
        y_{n+1} &= bx_n,
    \end{cases}
\end{align}
where $\lambda=(a,b) \in \mathbb{R}^2$ are parameters. Here we restrict our analysis to parameters in the set $\Gamma \subset \mathbb{R}^2$, with
\begin{equation}\label{eq.Gamma}
    \Gamma = \left\{(a, b): a \in \left(\frac{-(1-b)^2}{4}, \frac{3(1-b)^2}{4} \right), |b| < 1, a \neq 0 \right\},
\end{equation}
because for $\lambda \in \Gamma$ the point 
\begin{equation}\label{eq.henonfixedpoint}
    (x^*,y^*) = \left(\frac{-(1-b)+\sqrt{(1-b)^2+4a}}{2a},  bx^*\right)
\end{equation} 
is a {\it stable} fixed point of Eq.~\eqref{eq:henon}. Figure~\ref{fig.henon1} plots the set $\Gamma$ (panel a) and the basin of attraction of the stable fixed point $(x^*,y^*)$ (panel b) for two (pairs of) parameters $\lambda \in \Gamma$. The numerical results confirm that both smooth and fractal basin boundaries are observed, depending on the parameters~\cite{grebogi_basin_1987}.

\begin{figure}[!t]
    \centering

    \begin{subfigure}{0.46\textwidth}
        \centering
        \caption{\justifying }
        \includegraphics[width=1\linewidth]{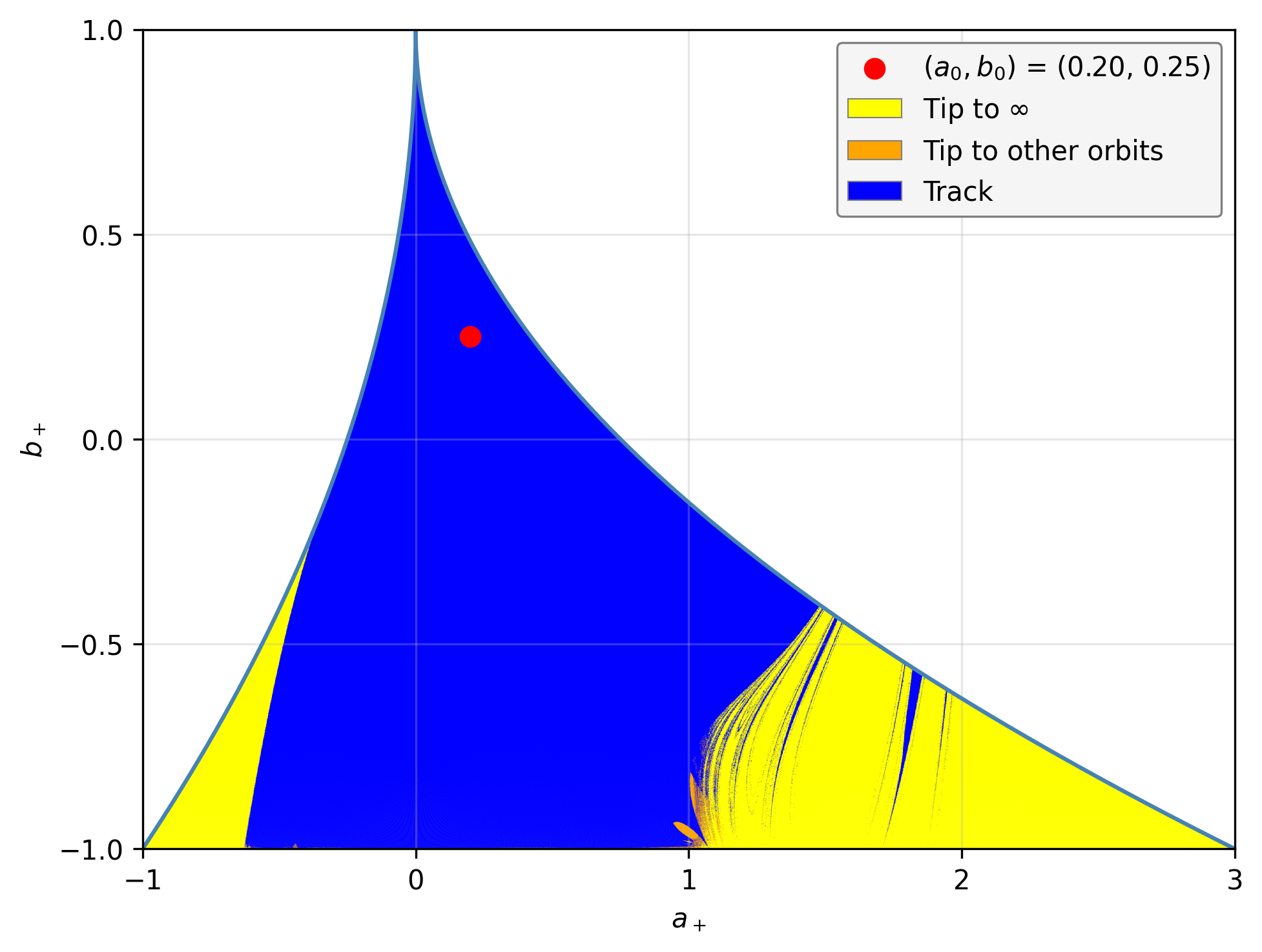}
    \end{subfigure}
    \hfill
    \begin{subfigure}{0.46\textwidth}
        \centering
        \caption{\justifying }
        \includegraphics[width=1\linewidth]{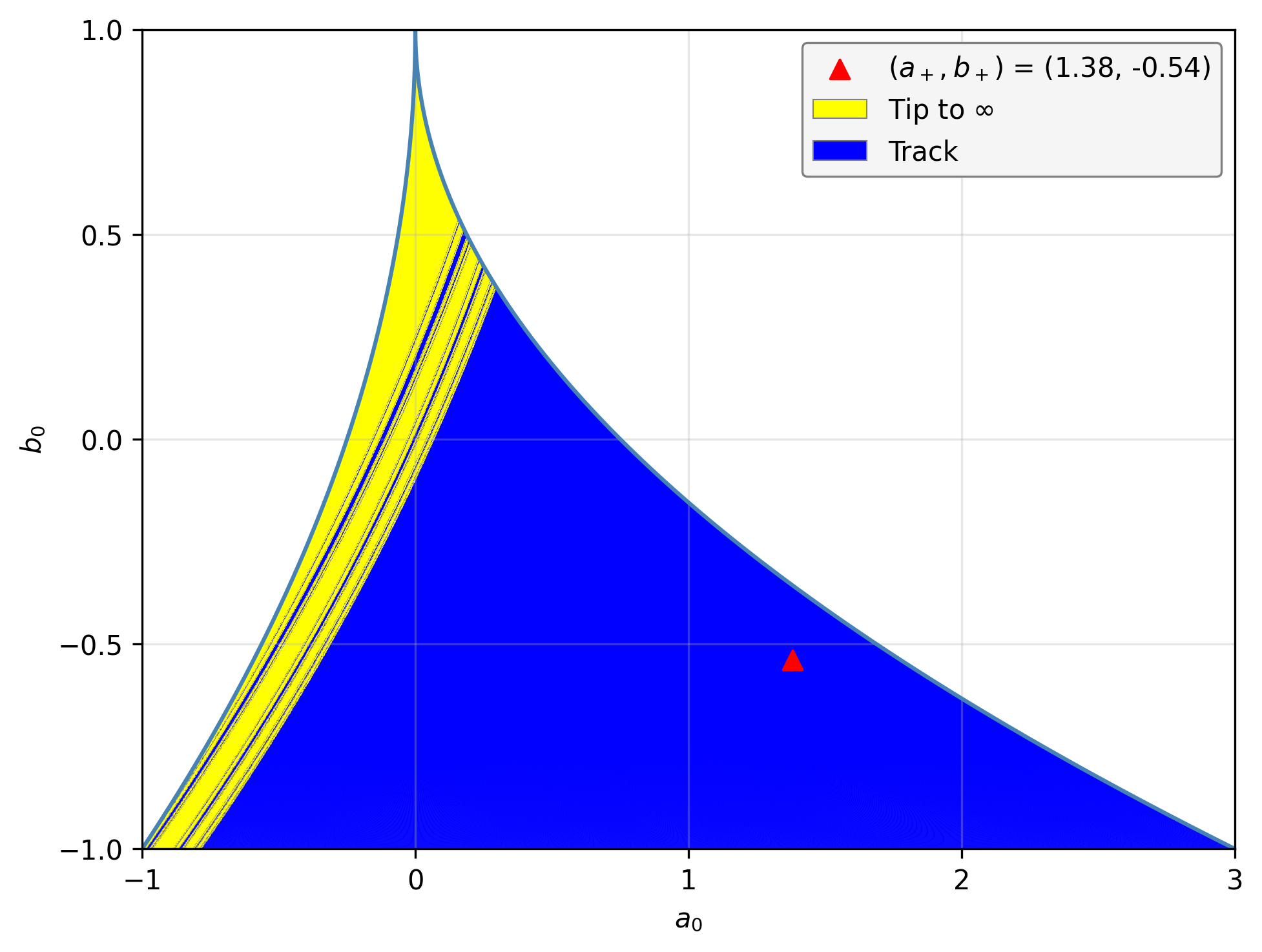}
    \end{subfigure}
    \caption{\justifying Fractal boundary between tracking and tipping regions in the parameter space $\Gamma$ of the H\'enon map. 
    Initial conditions at the fixed point $(x_0, y_0)= (x^*_0,y^*_0)$ of $\lambda_0=(a_0,b_0)$ are iterated using Eq.~(\ref{eq:henon}) with $\lambda=\lambda_+=(a_+,b_+)$.
    (a) Dependence of $\phi$ on $(a_+,b_+)$ for a fixed $(a_0,b_0)=(0.20,0.25)$.  (b) Dependence of $\phi$ on $(a_0,b_0)$ for a fixed $(a_+,b_+)=(1.38,-0.54)$. The fixed values of $\lambda_0$ and $\lambda_+$ (red symbols, see legend) are the same used in Fig.~\ref{fig.henon1}. Parameters $\lambda_0$ and $\lambda_+$ showing $\phi=$ tip will display R-tipping with at least one critical tipping rate $r_c$. In both plots, blue indicates tracking the fixed point, yellow indicates tipping to infinity, and orange indicates tipping to attractors different from the fixed point (e.g., periodic orbits).}
    \label{fig:HenonRtipRegion}\label{fig.henon2}
\end{figure}

To observe R-tipping in the H\'enon map using a stable path $\Lambda(s)$, we vary the parameter pair $(a, b)$ from a starting pair $\lambda_0 = (a_0, b_0)\in \Gamma$ to an end pair $\lambda_+ = (a_+, b_+) \in \Gamma$. The simplest choice of path, which we will use in our analysis for finite $r$, is a linear path between these points parameterized by
\begin{eqnarray}\label{eq:hennonDrift}
    \Lambda_B(s) = 
    \begin{cases}
     \lambda_0 & \text{ for } s < 0, \\
    \lambda_0 + (\lambda_+ - \lambda_0)\tanh(s)  & \text{ for } s \geq 0 ,
    \end{cases}
\end{eqnarray}
where $\lambda=(a,b)$ and $s=rn$. In Eq.~(\ref{eq:hennonDrift}), we restrict the change to $s>0$ (i.e., we take $\lambda_-=\lambda_0$) -- contrary to the approach taken in the first example~\eqref{eq:tanh_drift}. The reason for this choice is twofold: first, we want to illustrate the validity of our results for different choices of paths $\Lambda(s)$; second, we want to simplify the numerical calculations\footnote{In particular, taking $\lambda_-=\lambda_0$ allows us to check the track/tip condition for $r\rightarrow \infty$ in Fig.~\ref{fig.henon2} for any two parameter pairs in $\Gamma$, even if the linear path in Eq.~(\ref{eq:hennonDrift}) leaves $\Gamma$ leading to an unstable path for $r<\infty$.}. Below we report numerical results of the tracking/tipping function~\eqref{eq.phi} based on the dynamics~\eqref{eq.map} for the map $F=F_B$ in Eq.~\eqref{eq:henon} with parameters evolving as in Eq.~\eqref{eq:hennonDrift} and initial condition given by the fixed point~\eqref{eq.henonfixedpoint} with $(a,b)= (a_0,b_0) \equiv \lambda_0$.

\begin{figure}[!t]
    \centering

    \begin{subfigure}{0.45\textwidth}
        \centering
        \caption{\justifying }
        \includegraphics[width=1\linewidth]{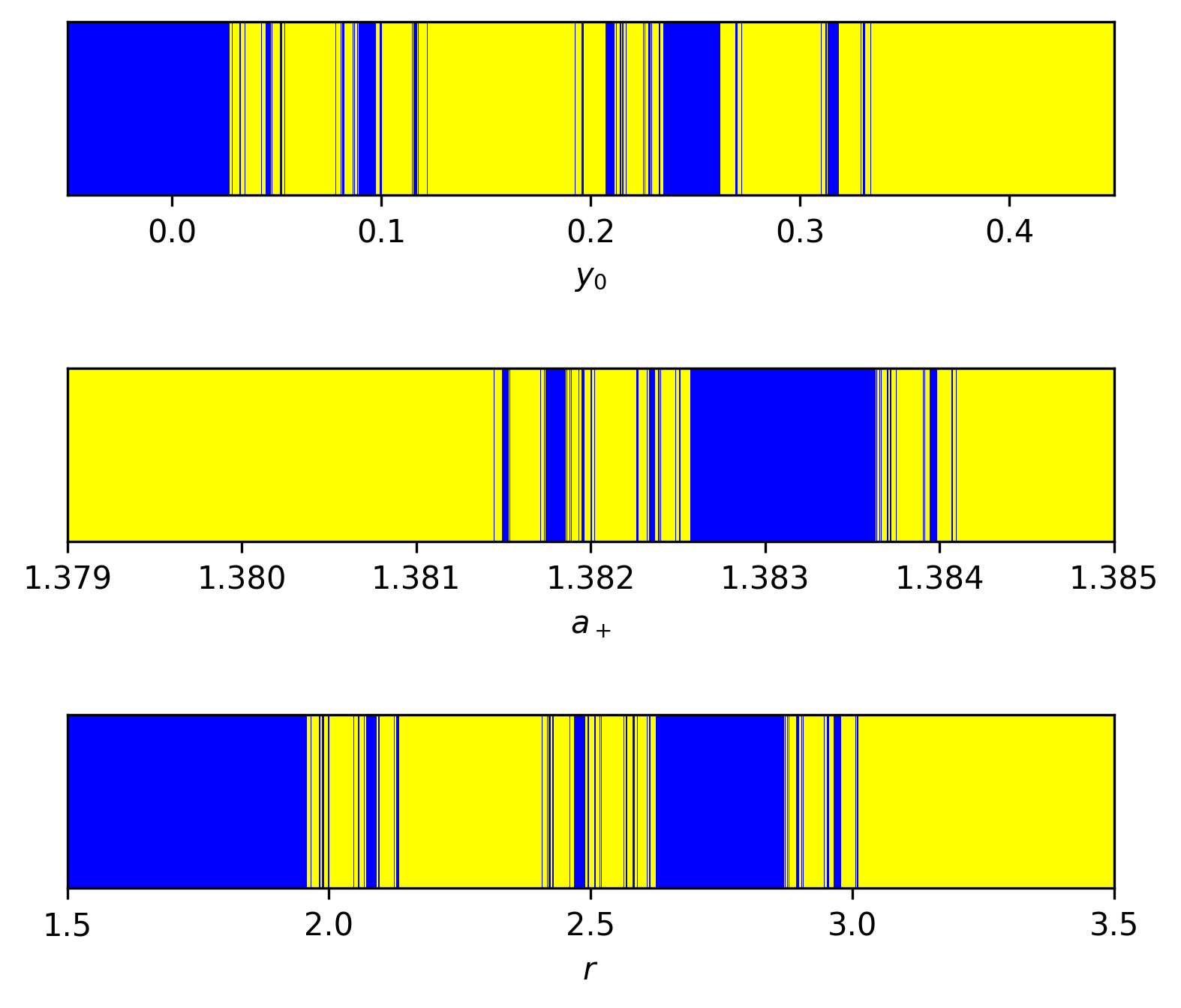}
    \end{subfigure}
    \hfill
    \begin{subfigure}{0.46\textwidth}
        \centering
        \caption{\justifying }
        \includegraphics[width=1\linewidth]{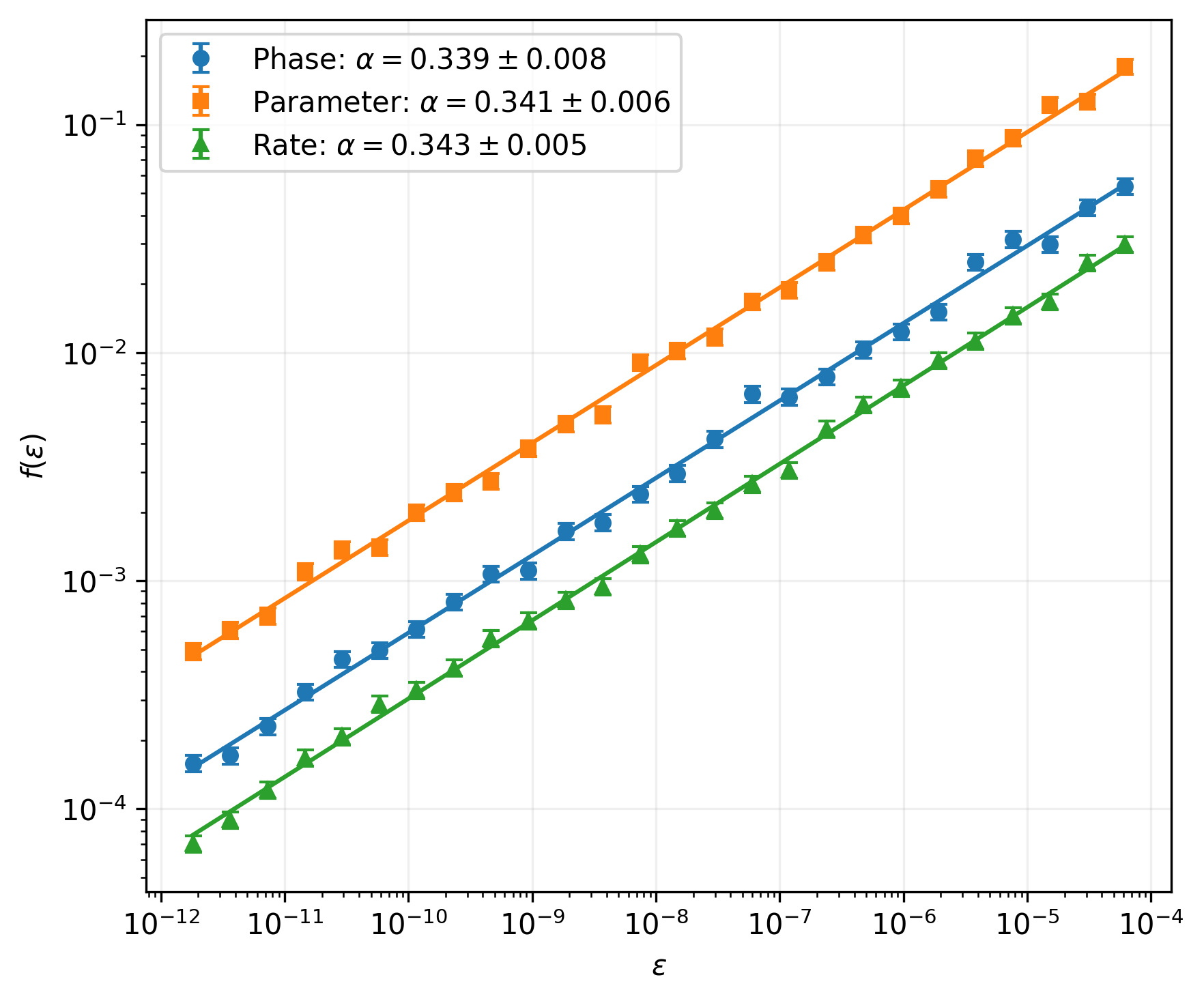}
    \end{subfigure}
    \caption{\justifying The three fractals in the H\'enon map~\eqref{eq:henon} have the same co-dimension~$\alpha$. (a) Plot of $\phi(r,x_0;\Lambda(s))$ in Eq.~(\ref{eq.phi}) -- with blue corresponding to $\phi=$ track and yellow to $\phi=$ tip (to infinity) --  as a function of different parameters (top to bottom): initial conditions $(x_0 =-0.2, y_0)$, corresponding to the basin-of-attraction of the fixed point for the end parameter $(a_+,b_+)=(1.38, -0.54)$ (the phase-space fractal); the end parameter value $(a_+,b_+=-0.54)$, for a fixed starting parameter $(a_0,b_0)=(0.20, 0.25)$ (the parameter-space fractal); rate $r$ using $(a_0,b_0)=(0.20, 0.25)$ and $(a_+,b_+)=(1.38, -0.54)$ (the rate-space fractal) under parameter path Eq.~\eqref{eq:hennonDrift}.
    (b) Estimation of the fractal co-dimension $\alpha$ of the track/tip boundary for the three fractals shown in panel (a), see Appendix~\ref{app.fractal} for details. }
    \label{fig:HenonFracBar3in1}\label{fig.henon3}
\end{figure}

We start our investigation considering whether R-tipping takes place for a given choice of $\lambda_0=(a_0,b_0)$ and $\lambda_+=(a_+,b_+)$. Considering that $\lambda$ changes in a stable parameter path inside~$\Gamma$, we know that for sufficiently small $r$ we have $\phi=$ track. Therefore, a value of $\phi=$ tip for $r\rightarrow \infty$ implies that at least one $r_c>0$ exists. Based on the results in Ref.~\cite{kiers_rate-induced_2020}, a sufficient condition is thus to check whether the fixed point~(\ref{eq.henonfixedpoint}) with $\lambda=\lambda_0$ is in the basin of attraction of the fixed point with $\lambda=\lambda_+$. This can be done by fixing either $\lambda_0$ or $\lambda_+$ and exhaustively exploring $\Gamma$. 
Figure~\ref{fig.henon2} shows the numerical results:
fixing $\lambda_0$, we obtain  -- Fig.~\ref{fig.henon2}(a) --- a parameter space $\Gamma$ with an intricate boundary between tracking and tipping, suggesting that smooth boundaries (on the bottom left) co-exist with fractal boundaries (bottom right); fixing $\lambda_+$ to a value in which the corresponding fixed point $(x^*,y^*)$ has a fractal basin boundary (Fig.~\ref{fig.henon1}(b), bottom right), we obtain -- Fig.~\ref{fig.henon2}(b) --- a separation of the parameter space $\Gamma$ that suggests a fractal boundary with a dimension between 1 and 2 (i.e., co-dimension $0<\alpha<1$).

We now focus on quantifying the fractality of the tracking/tipping boundaries. We focus on three cases: the basin of attraction of the fixed point at $\lambda_+$ (the phase-space fractal, shown in Fig.~\ref{fig.henon1}(b), right); the parameter space $\Gamma$ varying $\lambda_+=(a_+,b_+)$  for $r\rightarrow \infty$ (the parameter-space fractal, shown in Fig.~\ref{fig.henon2}(a)); and the dependence on the rate $r$ for fixed $\lambda_0$ and $\lambda_+$, in which cases the boundaries correspond to critical values $r_c$ (the rate-space fractal).  The numerical results are reported in Fig.~\ref{fig:HenonFracBar3in1} and strongly suggest that the three boundaries are fractal with the same co-dimension $\alpha$.  For simplicity, the phase-space fractal and the parameter-space fractal have their $\alpha$ computed from the 1D-line segment shown in Fig.~\ref{fig.henon3}(a). Importantly, the rate-space fractal is computed choosing a value of $\lambda_+$ for which $\phi=$ tip when $r\rightarrow \infty$, which we argued above ensures the existence of at least one critical $r_c$. In fact, our results suggest that there infinitely many $r_c$'s, distributed in a fractal set with dimension $D=1-\alpha \approx 0.657 \pm 0.005$. This is obtained for a path $\Lambda(s)$ that finishes at a parameter $\lambda_+=(a_+,b_+)$ for which $x^*$ has a fractal basin boundary.  Instead, if a $\lambda_+$ with a smooth boundary is chosen (see Figs.~\ref{fig.henon1}-\ref{fig.henon2}), a smooth transition in the $\phi$ dependence on $\lambda_+$ and $r$ are observed, as shown in Appendix~\ref{app.henonsmooth}.

\subsection{Forced Pendulum}

\begin{figure*}[!t]
    \centering
   \begin{subfigure}[t]{0.33\textwidth}
        \centering
        \caption{\justifying}
        \vspace{0pt}
        \includegraphics[width=\linewidth]{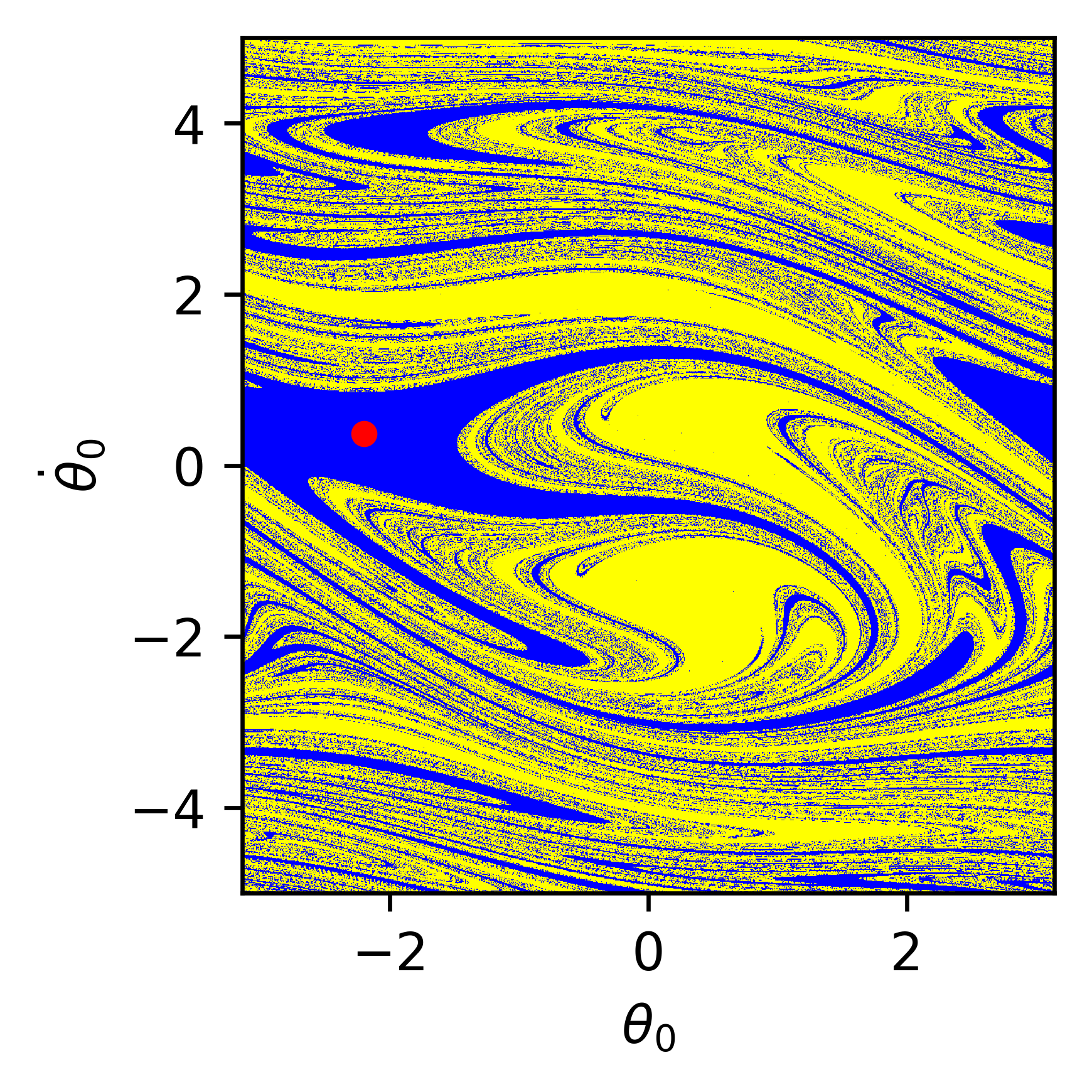}
        \label{fig:fractal1_full}
    \end{subfigure}
    \hfill
    \begin{subfigure}[t]{0.32\textwidth}
        \centering
        \caption{\justifying}
        \vspace{0pt}

        \begin{minipage}[t][\linewidth][t]{\linewidth}
            \centering
            \hspace{-1mm}
            \includegraphics[width=0.93\linewidth]{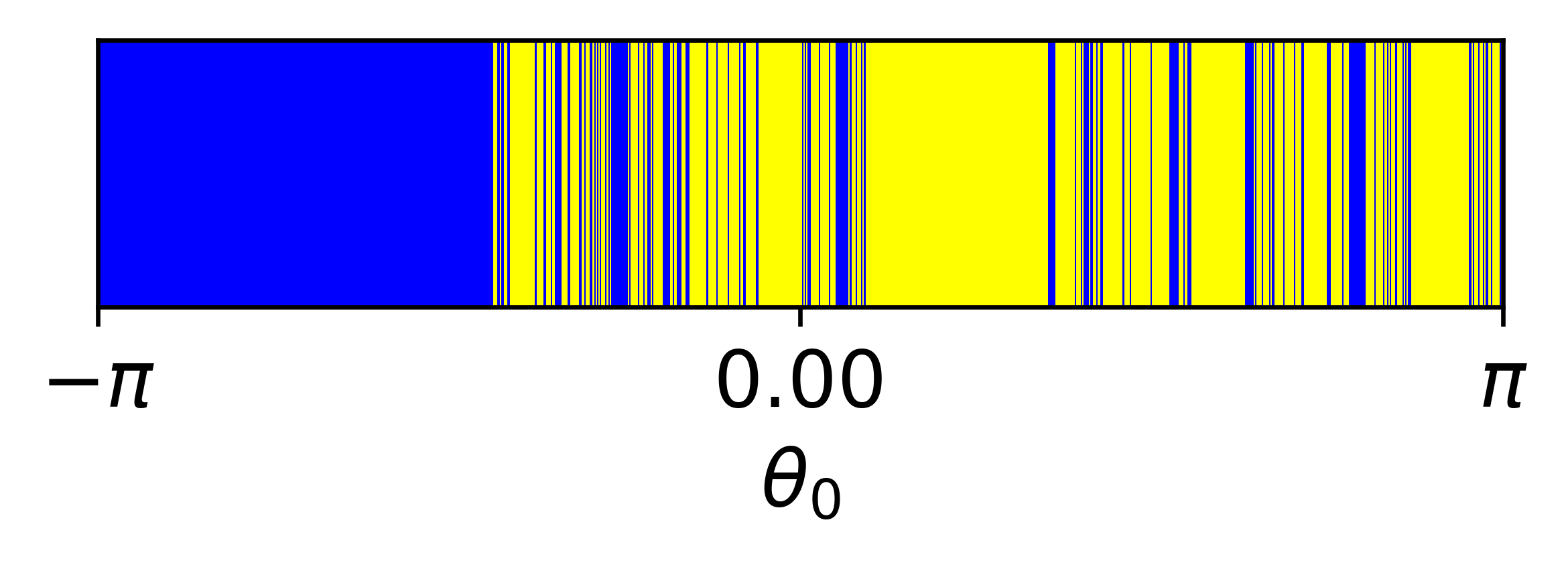}\vfill
            \includegraphics[width=\linewidth]{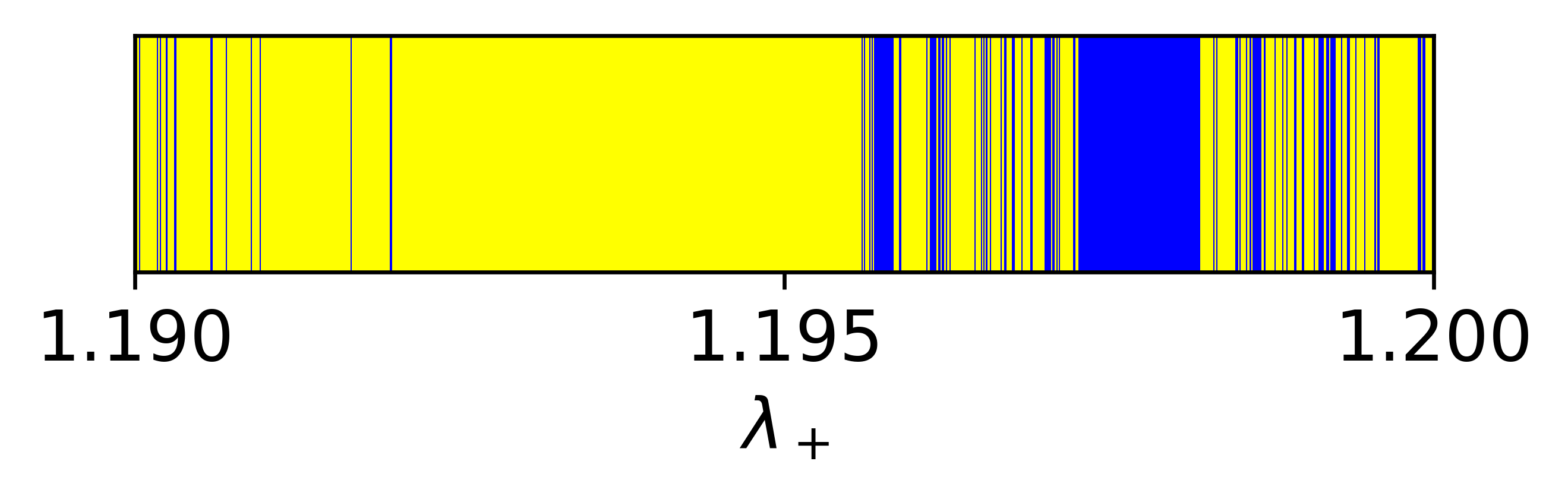} 
            \hspace{2mm}
            \includegraphics[width=0.98\linewidth]{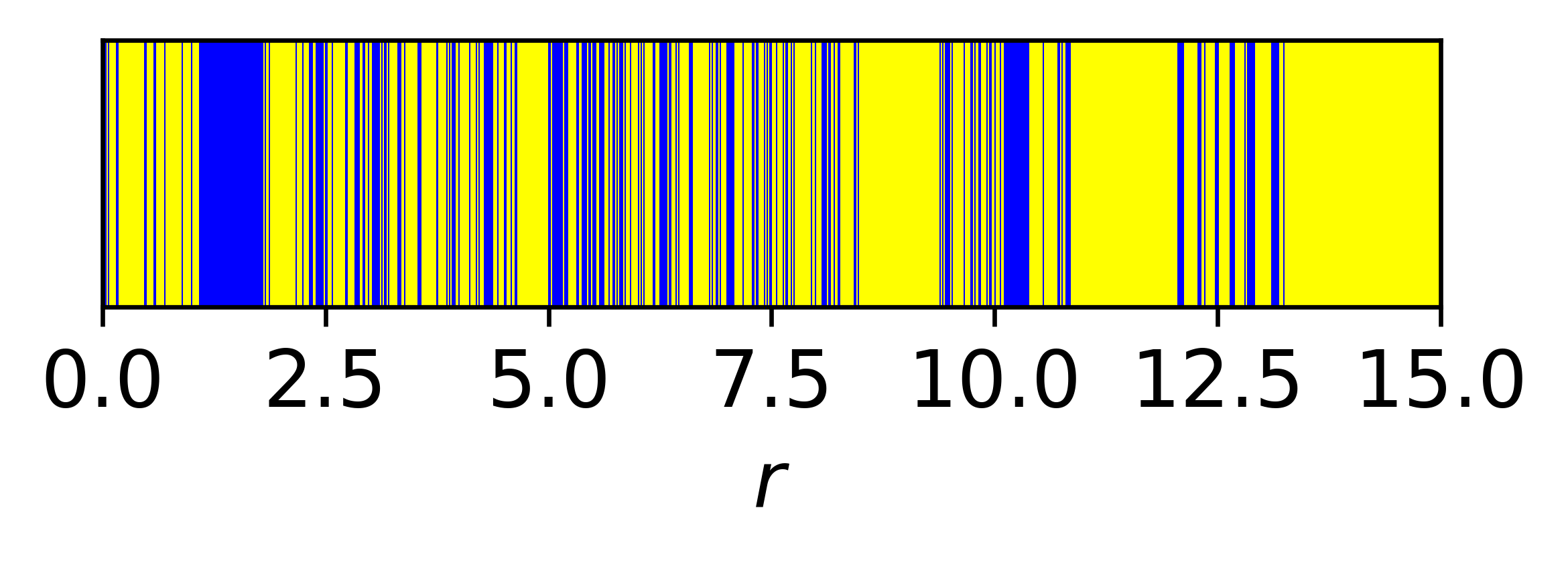}
        \end{minipage}

        \label{fig:fractals_pen}
    \end{subfigure}
    \hfill
    \begin{subfigure}[t]{0.3\textwidth}
        \centering
        \caption{\justifying}
        \vspace{0pt}
        \includegraphics[width=\linewidth]{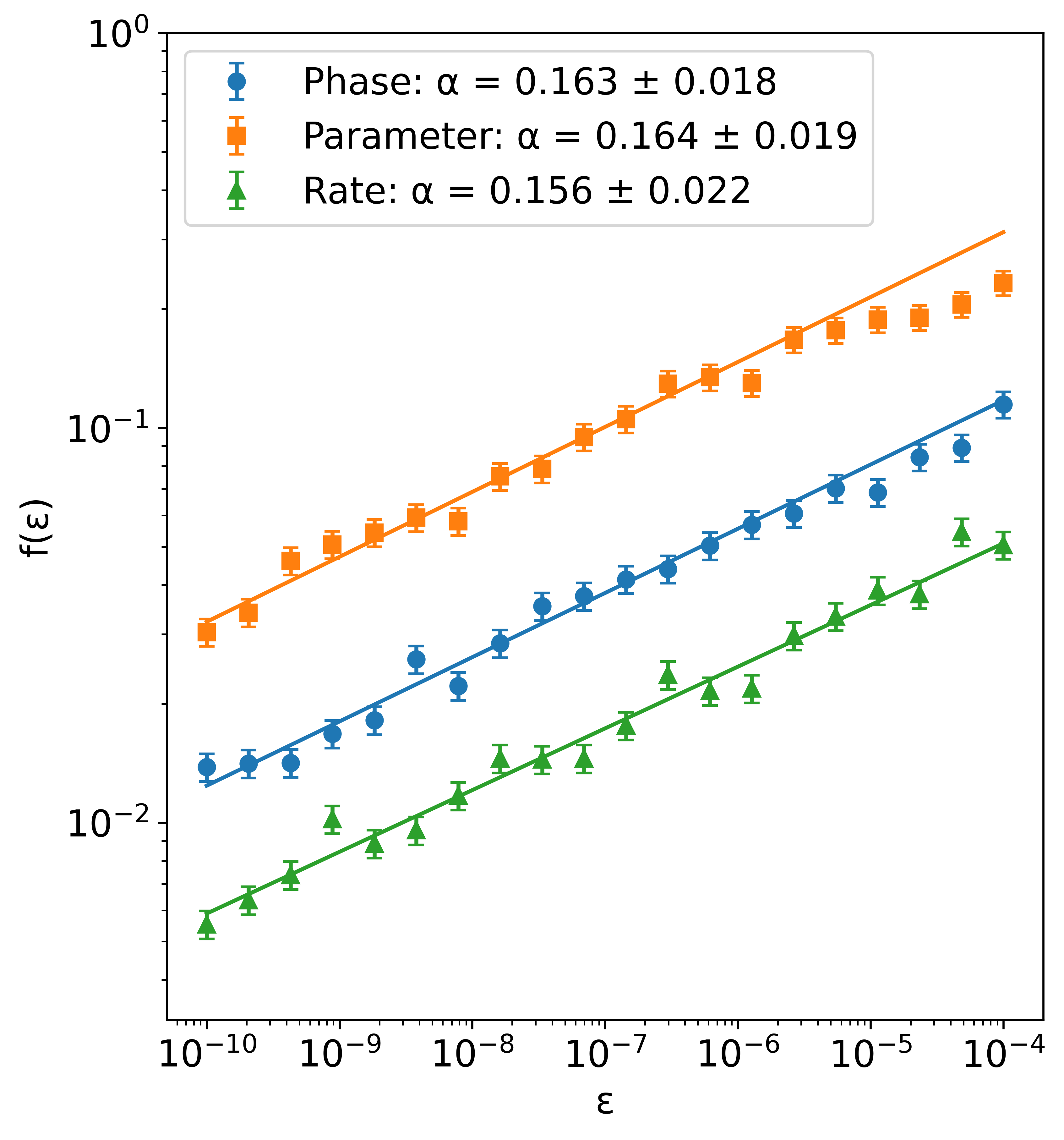}
        \label{fig:fits_pen}
    \end{subfigure}
    
        \caption{\justifying The three fractals in the forced pendulum~\eqref{eq.theta} have the same co-dimension~$\alpha$. (a) phase-space at the Poincar\'e surface of section showing the basin of attraction (in blue) of the fixed point $(\theta^*,\dot{\theta}^*) \approx (-2.20, 0.37)$ (red circle) for $\lambda=1.195$. All other final states (periodic or chaotic) are classified as tipped (in yellow). (b) Plot of the function $\phi(r,x_0;\Lambda(s))$ in Eq.~(\ref{eq.phi}) as a function of different variables, with blue corresponding to $\phi=$track and yellow to $\phi=$tip (to any final state). From top to bottom, the panels show the dependence of $\phi$ on the initial conditions $x_0 = (\theta, \dot{\theta}=0)$ (with $r=\lambda_+=0$, the phase-space fractal),
        range of parameter variation $\lambda_+$ used in $\Lambda(s)$ (with $r\rightarrow \infty$, the parameter-space fractal), and the rate of change $r$ (with $\Delta \lambda = 1.195$, the rate-space fractal, $r_{c1} \approx 0.040$ and $r_{c2} \approx 13.246$).  (c) Estimation of the fractal co-dimension $\alpha$ of the track/tip boundary for the three fractals shown in panel (b), see Appendix~\ref{app.fractal} for details. All three co-dimensions $\alpha$ were computed from the 1D-line segments shown in panel (b).  }
        \label{fig:fractal_fits_pen}
\end{figure*}

Our third example is a physical system: the continuous-time $t$ dynamics of a damped forced pendulum\cite{grebogi_basin_1987,Kaszas2016}, defined by the second order differential equation
\begin{equation}\label{eq.theta}
    \ddot{\theta} + 0.1 \dot{\theta} +\sin(\theta) = \lambda \cos(t),
\end{equation}
where $\theta \in[0,2\pi]$ is the angle position of the pendulum, $\dot{\theta}\equiv d\theta/dt$  is the angular velocity, and $\lambda$ is the intensity of the forcing (the only free parameter). The other physical parameters are fixed: damping strength to $0.1$ and both the natural and forcing frequencies to $1$. We use a Poincar\'e surface of section obtained every period $2\pi$ of the forcing, i.e., for $t= 2\pi n$ with $n\in \mathbb{Z}$, which induces a discrete-time map $F_C$ -- as in Eq.~(\ref{eq.f}) -- that maps $(\dot{\theta},\theta)_n$ to $(\dot{\theta},\theta)_{n+1}$. In practice, $F_3$ is computed by integrating numerically Eq.~\eqref{eq.theta}~\cite{wang_fractal_tipping_2026}. For sufficiently small forcing strength $\lambda$, the oscillation with period one is a stable periodic orbit of the flow and thus a stable fixed point $x^*=(\dot{\theta}^*,\theta^*)$ of the map $F_C$.  As investigated in the seminal paper~\cite{grebogi_basin_1987}, at values $\lambda \gtrapprox 1$ the basin of attraction of this fixed point has a fractal boundary. 

Here we use this paradigmatic example of fractal basin-boundary to test the R-tipping fractal phenomenology observed in the previous examples. As the parameter-change protocol, we use a linear ramp from $\lambda=0$ to $\lambda=\lambda_+$ as
\begin{eqnarray}\label{eq:tanh_drift2}
    \Lambda_C(s) = 
    \begin{cases}
      0 & \text{ for } $s < 0$, \\
        s  & \text{ for } 0 \le s \le \lambda_+, \\
        \lambda_+ & \text{ for } s > \lambda_+, \\
    \end{cases}
    \end{eqnarray}
with $\lambda_+=1.195$ (in which a fractal basin boundary is present) and $s=rt$ in continuous time $t$, i.e., $\dot{s}=t$. This choice, which is different from  the discrete-time definition in Eq.~(\ref{eq.map}), is more natural considering the physical picture of this system (variable forcing of a pendulum, see also Ref.~\cite{Kaszas2016} for a similar choice) and it illustrates that our results are valid for different choices of protocols $\Lambda(s)$. The numerical results reported in Fig.~\ref{fig:fractal_fits_pen} show a fractal dependence on the tipping parameters $\lambda_+$ and $r$ with the same fractal dimension of the basin boundary, confirming the observations obtained in the two previous examples.

\section{Theoretical interpretation}\label{sec:theory}\label{sec.theory}

The numerical experiments above show that, in all three examples, fractals appear in the dependence of the tipping function $\phi$ on the rate $r$  and amplitude $\Delta \lambda$ (or $\lambda_+$) of parameter change. More precisely, the infinitely many values in which $\phi$ changes outcome are fractaly distributed with a fractal co-dimension that is equal to the co-dimension of the fractal basin boundary of the frozen system at the asymptotic parameter $\lambda=\lambda_+$. In this section, we provide a general theoretical explanation for these observations and discuss (sufficient) conditions for the appearance of fractals in R-tipping through the mechanism described below.

Our starting point is an autonomous dynamical system~(\ref{eq.f}) which shows multistability and in which the attractor in focus (i.e., the fixed point $x^*$) at the parameter  $\lambda=\lambda_+$ has a fractal basin boundary, i.e., a fractal boundary in the dependence of $\phi(x_0,r=0; \Lambda(s)=\lambda_+)$ on the initial condition $x_0 \in \Omega \subseteq \mathbb{R}^d$. 
The existence of such fractal basin boundaries is well known in a variety of non-linear dynamical systems~\cite{McDondald1985,grebogi_basin_1987,aguirre_fractal_2009}. 
In the typical transient chaos picture~\cite{lai_transient_2011}, the boundary is composed by the stable manifold $W^S$ of the chaotic (hyperbolic) saddle that controls long-term transients.  If the dimension of this manifold $D(W^S)$ satisfies $d-1 < D(W^S) < d$, then $W^S$ divides the $d$-dimensional phase-space into regions that may converge to different attractors and therefore $W^S$ may act as a fractal basin boundary~\cite{grebogi_basin_1987}. We thus write:
\begin{quote}
    {\bf Condition 1:} the existence of a fractal saddle in the phase-space $x\in \Omega \subseteq \mathbb{R}^d$ of $F$ at  $\lambda=\lambda_+$ with a stable manifold $W^S$ with dimension $D(W^S)$ such that the co-dimension $\alpha(W^S) = d - D(W^S)$ is
    $$0<\alpha(W^S) <1$$
    and $W^S$ acts as the boundary of the basin of attraction of the (fixed-point) attractor $x^*(\lambda_+)$.
\end{quote}

The uncertainty exponent $\alpha_p$ of the {\bf p}hase-space fractal computed in the previous section (Sec.~\ref{sec.examples}) provides an estimation of the co-dimension of the basin boundary and thus of the stable manifold of the fractal saddle, see Appendix~\ref{app.fractal} for details. Therefore, we have $\alpha(W^S)=\alpha_p$. The next two conditions ensure that this fractal object induces the appearance of fractals in the two natural settings of R-tipping,  the parameter- and rate-space fractals, involving the dependence of $\phi$ on $\Lambda(s)$ and $r$, respectively. 

We first focus on the conditions for the existence of R-tipping. The dynamical system $F$ at $\lambda_+$ is susceptible to rate-induced tipping if $x^*$ and its basin of attraction are sensitive to changes in $\lambda$~\cite{ritchie_rate-induced_2023}. In line with the sufficient conditions derived in Ref.~\cite{kiers_rate-induced_2020}, we consider:
\begin{quote}
    {\bf Condition 2:} the parameter path $\Lambda(s)$ must be such that $x^*(\lambda_0)$ (i.e., the attractor at the starting parameter $\lambda_0$) is outside the basin of attraction of $x^*(\lambda_+)$ (i.e., the attractor at the final parameter $\lambda_+$).
\end{quote}
More generally, a weaker condition that can still lead to tipping is to require $x^*$ to leave the basin of attraction at least once in the path $\Lambda(s)$, i.e., there are $s_a,s_b$, with $s_a<s_b$, such that $x^*(\Lambda(s_a))$ is {\it not} in the basin of attraction of the attractor at $s_b$. These conditions are closely related to the notion of (forward) basin instability~\cite{ashwin_parameter_2017, Okeeffe2020, kiers_rate-induced_2020}. 

The last requirement for the appearance of fractals in R-tipping is to connect variations in parameters (e.g., $\lambda_+,r$) of the dynamics with variations in the trajectories in the phase-space:
\begin{quote}
    {\bf Condition 3:} the autonomous dynamics $F$ and parameter path $\Lambda(s)$ must be such that continuous variations in the parameter $\lambda$ along $\Lambda(s)$ for a fixed initial condition $x_0$ leads to a continuous curve $F^{(n)}(x_0;\lambda)$ in the phase-space $\Omega$ for $n>0$.
\end{quote}
This last condition is satisfied in the examples above due to the continuity of $F(x;\lambda)$ and $\Lambda(s)$.

\begin{figure}[!t]
    \centering
    \begin{subfigure}{0.5\textwidth}
        \centering
        \caption{\justifying }
        \includegraphics[width=0.48\linewidth]{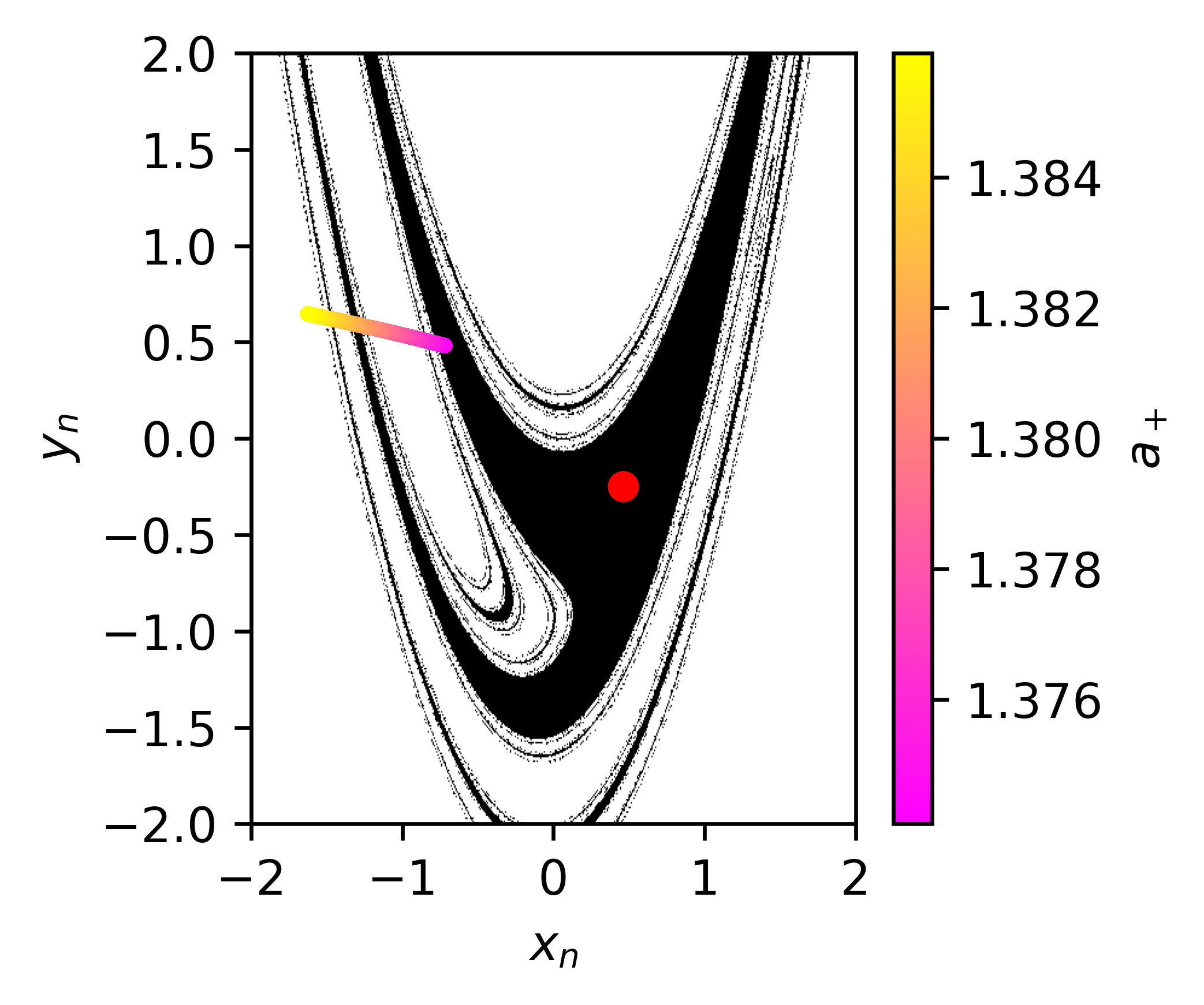}
        \includegraphics[width=0.48\linewidth]{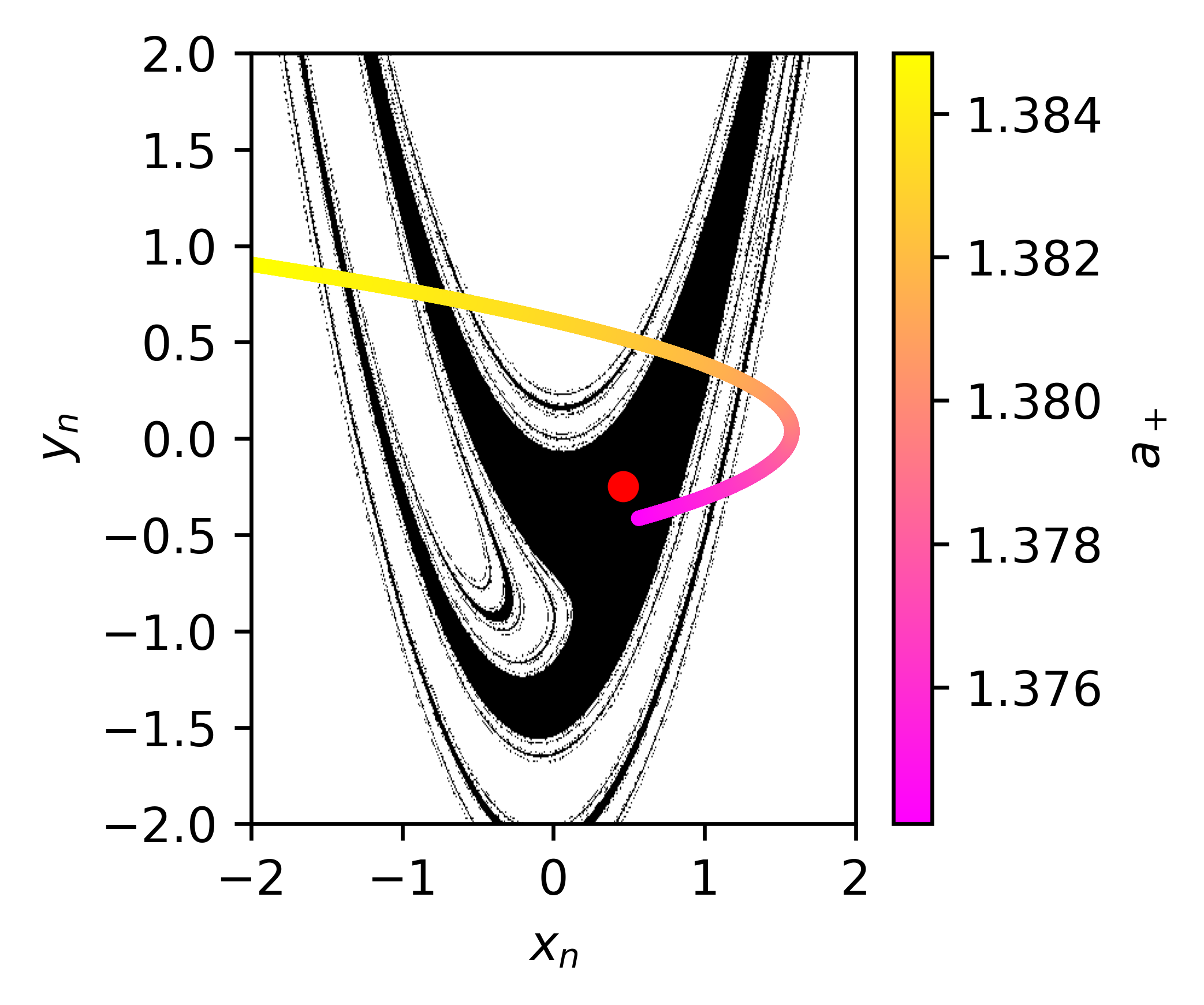}
    \end{subfigure}
    \hfill
    \begin{subfigure}{0.5\textwidth}
        \centering
        \caption{\justifying }
        \includegraphics[width=0.48\linewidth]{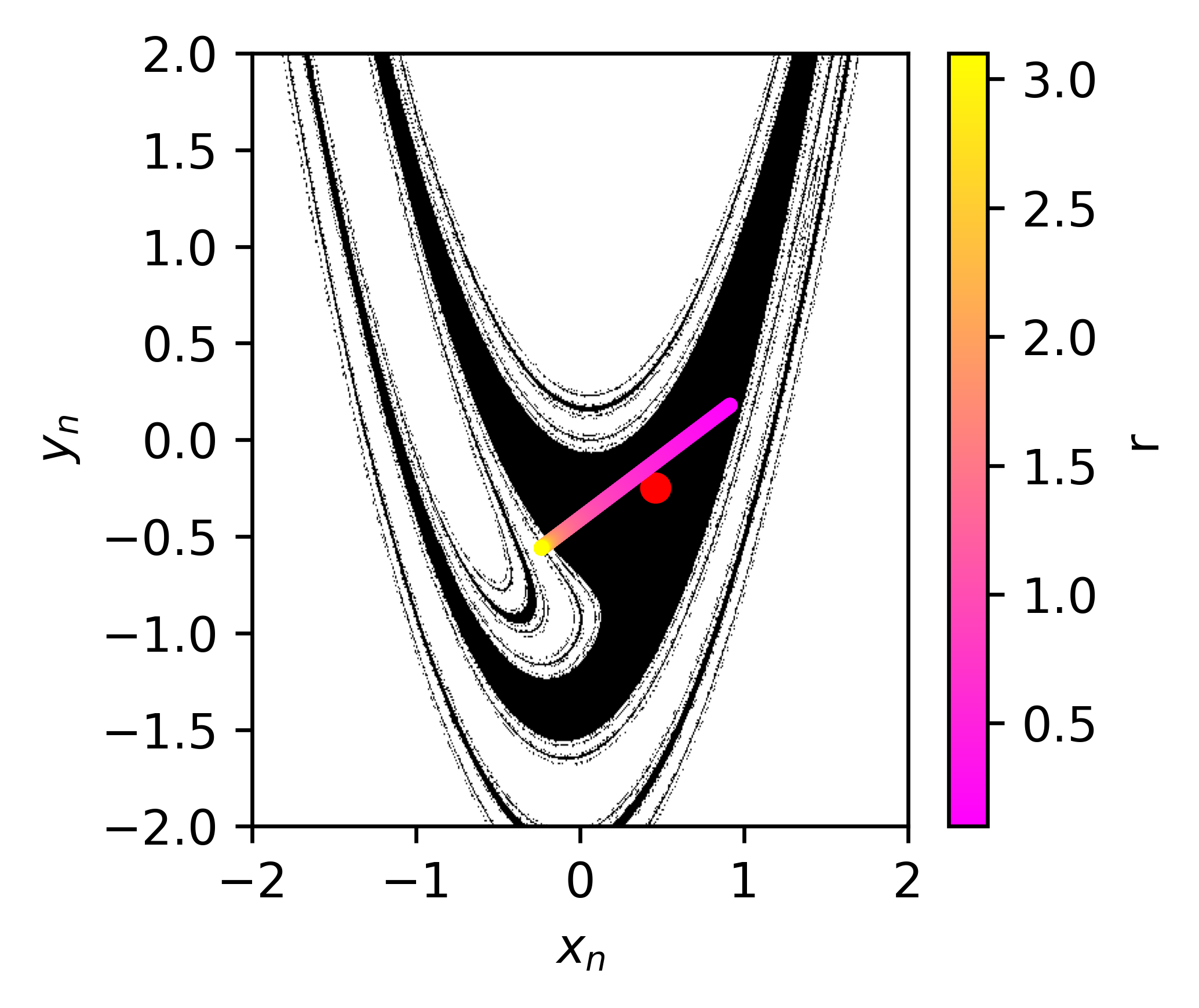}
        \includegraphics[width=0.48\linewidth]{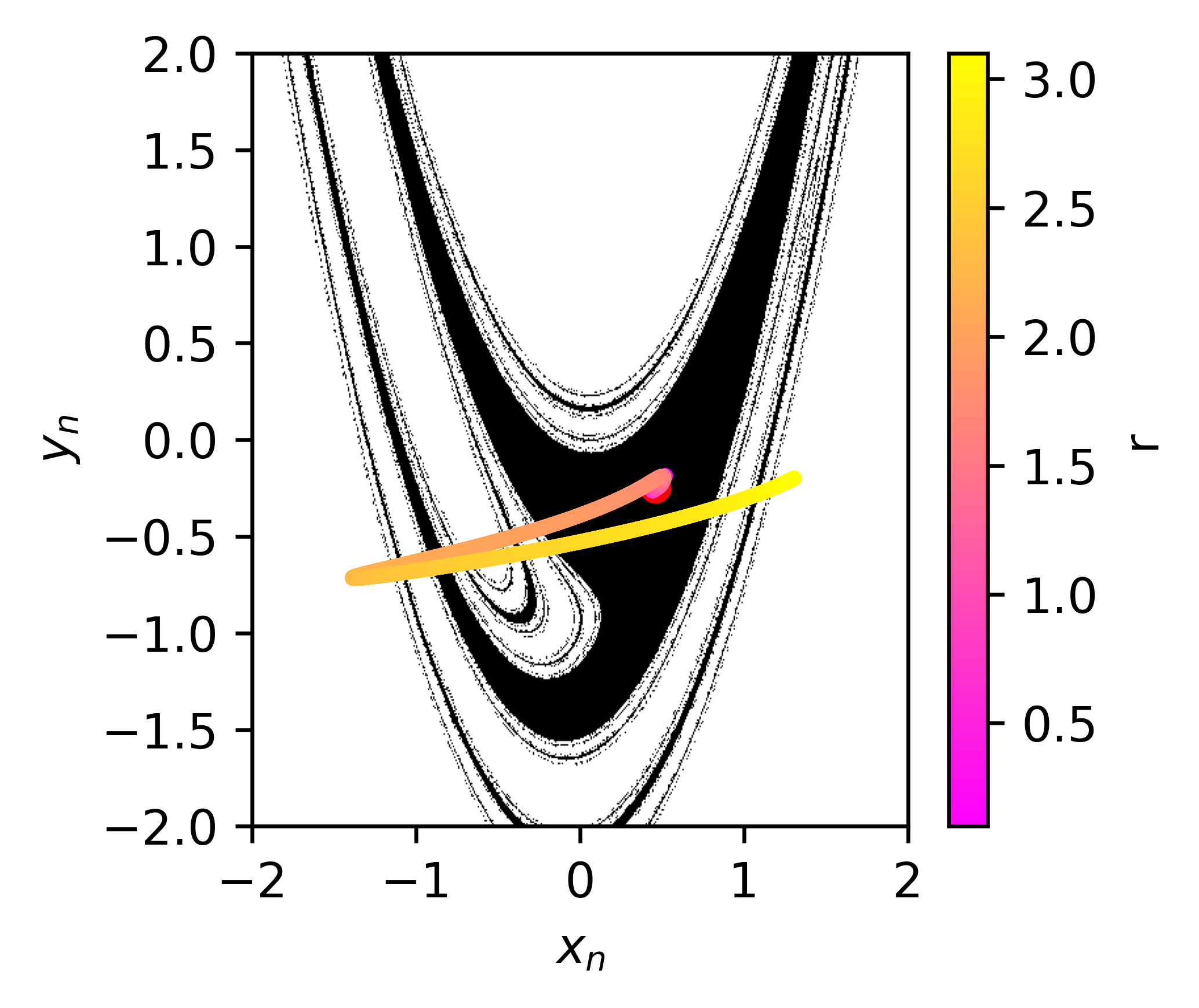}
    \end{subfigure}
    \caption{\justifying Continuous parameter variations lead to continuous curves in the phase-space of the H\'enon map $F_B(x,\lambda)$ in Eq.~(\ref{eq:henon}). The plots show curves in phase-space (in color) corresponding to $F_B^{(n)}(x^*, \lambda)$, with fixed initial condition $x^*$ (the fixed point at parameter $\lambda_0=(a_0,b_0)=(0.20,0.25)$) and continuously varying parameters $\lambda=(a,b)$. The background shows the fixed point (in red) and its basin of attraction (in black) for $\lambda = (a_+, b_+)=(1.38,-0.54)$, as in the right panel of Fig.~\ref{fig.henon1}(b). (a) Parameter change $\lambda = (a_0 + (a_+ - a_0)\eta, b_0 + (b_+ - b_0)\eta)$, with $\eta\in\mathbb{R}$ parameterizing the curve. Left: $n=8$; Right: $n=10$. We choose $\eta \in [0.995,1.005]$ so that $\lambda$ is close to $(a_+,b_+)$. (b) Parameter change $\lambda = \Lambda(s)$ defined in Eq.~\eqref{eq:hennonDrift} and $s=rn$, with $r \geq 0$ parameterizing the curve. Left: $n=1$; Right: $n=10$. For large $n$, the intersection of the curves with the basin boundary will correspond to the parameter-space and rate-space fractals shown in Fig.~\ref{fig.henon3}.}
    \label{fig:phase_cuts}
\end{figure}

We now discuss how these conditions lead to the appearance of fractals in R-tipping. Condition 1 ensures that there is a fractal basin boundary in the phase-space $\Omega$ (the phase-space fractal) at the asymptotic parameter $\lambda_+=\lim_{s\rightarrow \infty} \Lambda(s)$. 
For sufficiently small $r$, and stable path $\Lambda(s)$, $\phi(r,x_0; \Lambda(s))=$ track because $x_0=x^*(s_0)$ (i.e., initial condition in the attractor). Condition 2 ensures that $\phi(r\rightarrow \infty, x_0; \Lambda(s))=$ tip, and therefore, that there is at least one critical value $r_c$.  For $r = r_c$, the trajectory $x(n)$ hesitates between tracking and tipping, approaching an edge state as $n\rightarrow \infty$. In this limit, $\lambda \rightarrow \lambda_+$ and thus the non-autonomous dynamics in Eq.~(\ref{eq.map}) becomes effectively autonomous as in Eq.~(\ref{eq.f}) with $\lambda=\lambda_+$. Thus, the edge state approached by the trajectory will typically be the chaotic saddle associated with the phase-space fractal\footnote{When the trajectory starts at $s\rightarrow -\infty$, this trajectory can be thought as the snapshot or pullback-attractor which is therefore at the stable manifold of a pullback saddle~\cite{lai_transient_2011} that converges to the chaotic saddle as $n\rightarrow \infty$. When the limit $s \rightarrow -\infty$ is taken in such a way that $s=-rn$, with $n\in \mathbb{N}$, we ensure that the temporal evolution includes $s=0$ and thus $\lambda$ passes through $\lambda(s=0)$ (as done in Ref.~\cite{kiers_rate-induced_2020}); if, instead, we consider $s=s_0-rn$ with arbitrary $s_0$, we observed that there is no unique limit of $\phi$ as $s \rightarrow -\infty$, i.e., the sensitivity on the choice of $s_0$ persists, similar to the initial-condition dependence observed in partial tipping~\cite{alkhayuon_rate-induced_2018,Lohmann2021}.}. Condition 3 ensures that small (continuous) variations in $\lambda_+$ around a boundary of $\phi(r,x_0;\Lambda(s))$ lead to slightly perturbed trajectories that create a curve at fixed $n>0$ in the phase-space $\Omega \subseteq \mathbb{R}^d$. Varying $r$ leads to a similar curve, as $\lambda=\Lambda(rn)$. These curves expand exponentially in $n$ (due to the positive Lyapunov exponent of the saddle) so that for sufficiently large $n$ it will generically intersect the phase-space fractal (because its co-dimension is $0<\alpha <1$ as specified in Condition 1). 
These intersections lead to changes in the outcome of $\phi(r,x_0;\Lambda(s))$ at different critical values of $\lambda_+$ or $r$, leading to the appearance of the parameter- and rate-space fractals (respectively) with the {\it same} fractal co-dimension as the phase-space fractal:
\begin{equation}\label{eq.alphas}
    \alpha_p=\alpha_\lambda=\alpha_r.
\end{equation}
The curves mentioned in Condition 3 and in the argument above are shown in Fig.~\ref{fig:phase_cuts} for the case of the H\'enon map. These numerical simulations confirm the growth of the curves in time, their intersection with the phase-space fractal, and thus the connection leading to Eq.~(\ref{eq.alphas}). 
The equality $\alpha_p=\alpha_\lambda$ between the fractals in the basin of attraction (the phase-space fractal) and in the (frozen) parameter space (the parameter-space fractal) has been observed and discussed in Refs.~\cite{Moon1984,McDondald1985,lai1994}.

In the argument above, we have neglected the dependence of $\alpha_p$ on $\lambda_+$. While a single $\lambda_+$ is used in the phase- and rate-space fractals, and thus in the estimation of $\alpha_\lambda$ and $\alpha_r$, the parameter-space fractal considers precisely the variation of $\phi$ on $\lambda_+$. Therefore, the equality involving $\alpha_\lambda$ in Eq.~\eqref{eq.alphas} requires a local interpretation of the parameter-space fractal and $\alpha_\lambda$ to be estimated locally around a boundary value in $\lambda_+$, i.e., the computation of the pointwise-fractal dimension around a critical value $\Delta \lambda_c = (\lambda_+ - \lambda_-)_c$. In examples B and C, this motivated the use of small intervals of $\lambda_+$ around the value of interest as line segments for the computation of $\alpha_\lambda$ (see Fig.~\ref{fig.henon3}(a) and Fig.~\ref{fig:fractal_fits_pen}(a), middle panels). More generally, $\alpha_\lambda$ is dominated by the region of $\lambda_+$ with largest dimension (smallest $\alpha$). 
The dimensions in Eq.~(\ref{eq.alphas}) are computed for $\varepsilon \rightarrow 0$, which implies chaotic trajectories $n\rightarrow \infty$ and thus $\lambda \rightarrow \lambda_+$. For finite temporal and spatial scales $\varepsilon$, we expect the effective\cite{motter_effective_2005} fractal dimension $\alpha_\lambda(\varepsilon)$ to follow $\alpha_p(\lambda(n))$.

\section{Discussion and conclusions}\label{sec.conclusion}

We have investigated a generic mechanism for the appearance of fractals in R-tipping.
We have shown that if the attractor of a dynamical systems $F$ at parameter $\lambda_+$ has a basin of attraction with a fractal boundary, parameter changes that converge to $\lambda_+$ have tracking-tipping transitions in an (uncountable) infinite number of critical rates $r_c$. Moreover, the set of such $r_c$'s has the same fractal co-dimension of the basin boundary and of the tracking/tipping boundary as a function of $\lambda_+$. We illustrated this mechanism through numerical simulations in three simple systems, subject to parameter changes as done in R-tipping settings~\cite{ashwin_tipping_2012, kiers_rate-induced_2020}
\footnote{Our definition of tracking in Eq.~(\ref{eq.phi}) considers only the asymptotic state, differently from other definitions~\cite{Feudel2023} which reserve tracking to trajectories that follow closely the frozen path of the attractor and use transient tipping or reversible tipping~\cite{wieczorek_rate-induced_2023} to describe trajectories that depart the vicinity of the frozen attractor but eventually settle in it. In our case, such departure takes place for the trajectories associated with the three fractals as they approach the fractal saddle at $\lambda_+$. In our first example, $x^{u,3}$ separates these trajectories from the tracking trajectories that remain close to the attractor.}. We also provided a theoretical interpretation and discussed general conditions under which the equality of the co-dimensions -- Eq.~(\ref{eq.alphas}) -- is expected to hold. 

 Connections between chaotic transients, fractal edge states, and fractal basin boundaries have been investigated in different systems showing tipping behavior~\cite{Kaszas2016,Lucarini2017,kaszas_tipping_2019,lohmann_risk_2021,Lohmann2021,ashwin_physical_2021, mehling_limits_2024,Lohmann2024}. Our contribution here is to show how these phase-space fractals lead to the parameter- and rate-space fractals. We have done this in the simplest case of a single initial condition at a fixed point attractor, which remains stable for all parameters $\lambda$. R-tipping has been studied also in more general settings, involving periodic~\cite{alkhayuon_rate-induced_2018} and (fractal) chaotic~\cite{Kaszas2016,kaszas_tipping_2019,ashwin_physical_2021,alkhayuon_weak_2020,lohmann_predictability_2024} attractors, and considering ensembles of initial conditions~\cite{kaszas_tipping_2019,lohmann_risk_2021, ashwin_physical_2021,lohmann_predictability_2024,Kaszas2016}. We expect that future studies will be able to expand and connect our results to these more general settings, for instance, to the observation that fractal basin boundaries (in the phase-space) lead to a non-monotonic dependence of the tipping probability (of an ensemble of trajectories) on the rate $r$~\cite{kaszas_tipping_2019}.

We have tested the validity of our results for different parameter-variation protocols $\Lambda(s)$. Still, all of them had a limiting case $\lambda_+$. A more challenging case is whether our results can persist when no such limit is observed (e.g., when the parameter grow or decay continuously). This generalization requires considering the dimension of the pullback or snapshot saddle~\cite{lai_transient_2011,janosi_overview_2024}. One such scenario can be analyzed in our first example ($d=1$ piecewise-linear map): a linear ramp $\Lambda(s)= s=rn$ leads to a single critical tipping rate $r_c$; however, a fractal is still observed if the ramp is initially fast ($d\Lambda(rn)/dn > r_c$) but slows down ($d\Lambda(rn)/dn < r_c$) for large $s$. This example shows that fractals can also appear in R-tipping for more general parameter-changing protocols $\Lambda(s)$, provided that convergence to the moving attractor $x_n \rightarrow x^*$ is possible for large $s$.

Our analysis in simple dynamical systems intends to contribute to the understanding of generic rate-induced tipping phenomena.
The ubiquity of fractal saddles and transient chaos in non-linear dynamics~\cite{lai_transient_2011,aguirre_fractal_2009} suggests that the fractals we observe can appear also in non-autonomous systems proposed as models of physical or biological systems. In fact, fractal edge states have been reported in climate models with tipping points~\cite{Lucarini2017,lohmann_risk_2021,Lohmann2024, ashwin_physical_2021, lohmann_predictability_2024}. Our results predict fractals in R-tipping with the same co-dimension $\alpha$ as the basin boundary fractal, as stated in Eq.~(\ref{eq.alphas}). In such cases, our results can have practical consequence, such as quantifying (through $\alpha$) the extreme sensitivity to small perturbations and extreme oscillations, as well as opportunities for intervention or control. We further expect that the full range of fractal phenomena present in transient chaos~\cite{aguirre_fractal_2009,lai_transient_2011,altmann_leaking_2013} also appear in the R-tipping setting. For instance, we expect a fractal boundary separating three different asymptotic states (as in Fig.~\ref{fig.henon2}a) and the Wada property~\cite{aguirre_fractal_2009} to appear in the dependence of $\phi$ on $r$.

\section*{Acknowledgments}

EGA thanks Hinke Osinga for helpful suggestions.

\appendix \label{sec.appendix}

\section{Complete expression of the modified tent map}\label{sec.appnedixMap1} 

A general expression for the extended tent map, as in Eqs.~(\ref{eq.example1}) and (\ref{eq.lambdaTent}), valid for more general parameters is
 $x_{n+1} = F_A(x_n)$ with
\begin{eqnarray*}\label{eq.example1.complete}
 F_A =
\begin{cases}
\mu \,\min\!\bigl(x_n-\lambda,\, 1-(x_n-\lambda)\bigr)+\lambda, & x_n-\lambda < s_{1}, \\
a_{1}\bigl(x_n-\lambda\bigr)-b_{1}+\lambda, & s_{1} \leq x_n-\lambda < s_{2}, \\
a_{2}\bigl(x_n-\lambda\bigr)-b_{2}+\lambda, & x_n-\lambda \geq s_{2},
\end{cases}
\end{eqnarray*}
with $\lambda, \mu>2, 0<a_2<1,$ and $1<s_2< 3\mu/4 -1/2$ as free parameters. 
We fix the remaining parameters $a_1,b_1,b_2,$ and $s_1$ using the same constraints discussed in the main text after Eq.~(\ref{eq.example1}), i.e., setting the escaping region, re-injection region, and the attracting region of the intermediate regime to be the same in size. This leads to

\begin{align}
a_1 &= \frac{e_2}{e_2 - e_1},\\
e_1 &= 1 + \tfrac{1}{3}(s_2 - 1),\\
e_2 &= 1 + \tfrac{2}{3}(s_2 - 1),\\
a_2 &= \tfrac{1}{4},\\
b_1 &= a_1 s_1 - \mu(1 - s_1),\\
b_2 &= b_1 + s_2 (a_2 - a_1),\\
s_1 &= \frac{a_1 e_1 + \mu}{a_1 + \mu},\\
s_2 &= \frac{\mu}{2}.
\end{align}
The case used in the main text -- Eq.~(\ref{eq.example1})  and Fig.~\ref{fig:modified_tent} -- is retrieved choosing $a_2=1/4$, $\mu=3, s_2 = \mu/2,$ and $\lambda=0$.

\section{Two-parameters analysis of the piece-wise linear map}\label{sec:two_params}

\begin{figure}[!t]
    \centering
        \includegraphics[width=0.8\linewidth]{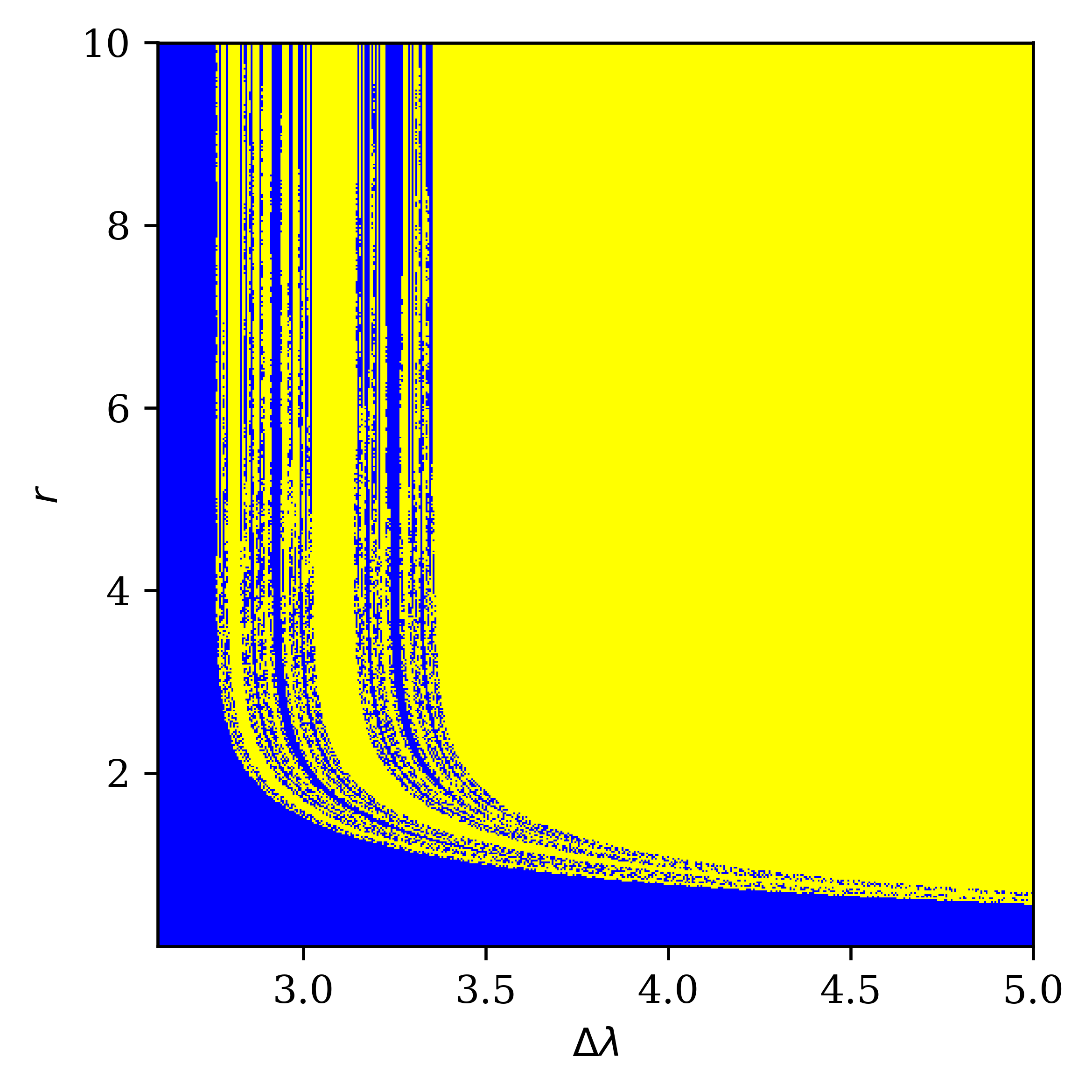}
    \caption{\justifying  Tipping and tracking in the piecewise-linear map~(\ref{eq.lambdaTent}) as a function of the two parameters $(\delta \lambda, r)$ used in $\Lambda(s)$ defined in Eq.~(\ref{eq:tanh_drift}). The plot shows the function $\phi(r,x_0;\Lambda(s))$ in Eq.~(\ref{eq.phi}), with blue corresponding to $\phi=$track and yellow to $\phi=$tip. The two lower stripes in Fig.~\ref{fig:combined}(a) correspond to $1-d$ cuts of the plot shown here: at $r\rightarrow \infty$ (for the parameter-space fractal) and at $\Delta \lambda = 2\lambda_+=4$  (for the rate-space fractal). Alternative choices lead to the same fractal-codimension $\alpha$.     }
    \label{fig:2d}
\end{figure}

Fig.~\ref{fig:2d} shows numerical simulations of the tracking/tipping outcome of the piecewise-linear map~(\ref{eq.lambdaTent}) as a function of two parameters. 

\begin{figure}[!t]
    \centering
    \begin{subfigure}{0.41\textwidth}
        \centering
        \caption{\justifying }
        \includegraphics[width=0.5\linewidth]{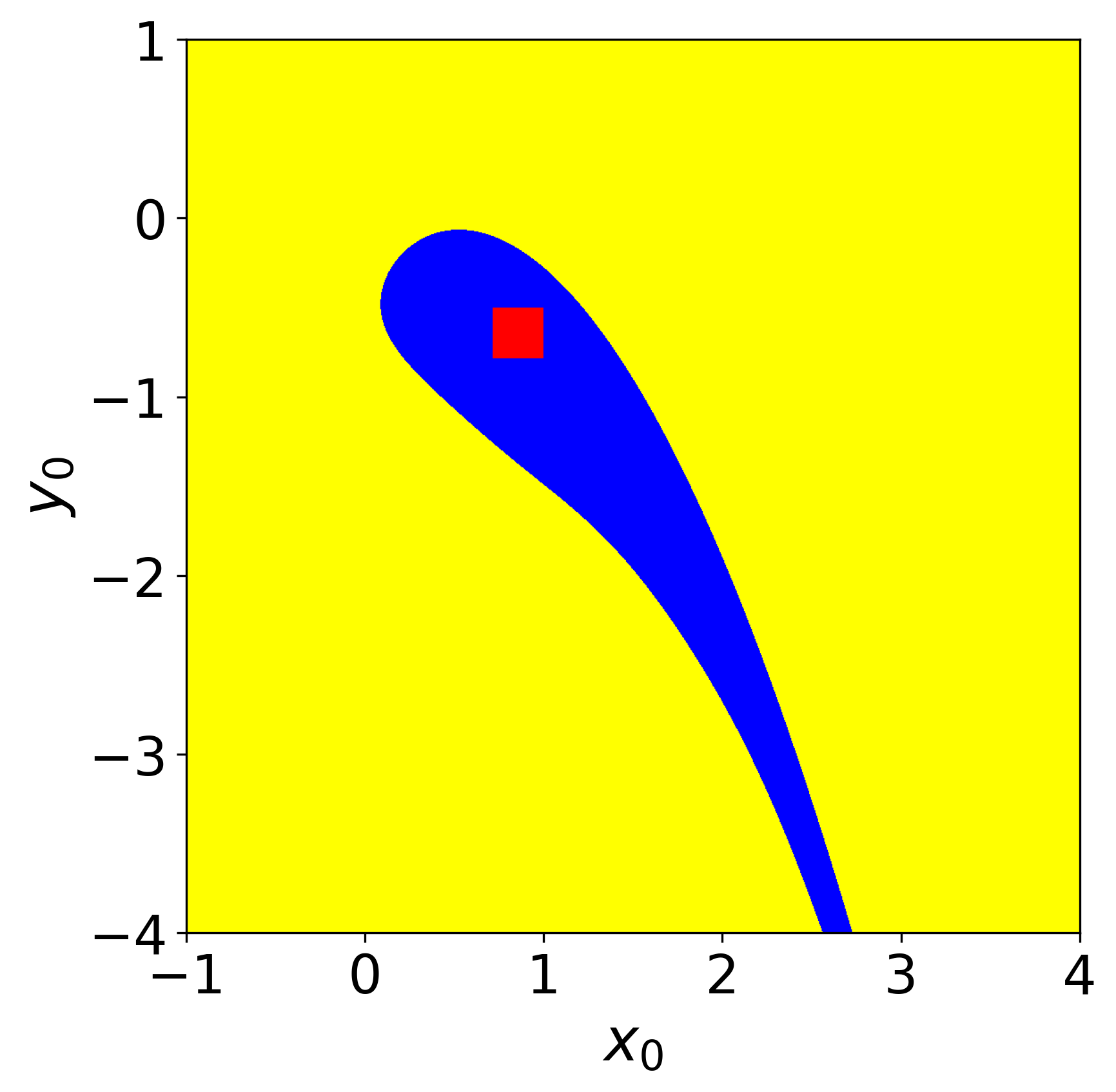}
    \end{subfigure}
    \hfill
    \begin{subfigure}{0.46\textwidth}
        \centering
        \caption{\justifying }
        \includegraphics[width=0.7\linewidth]{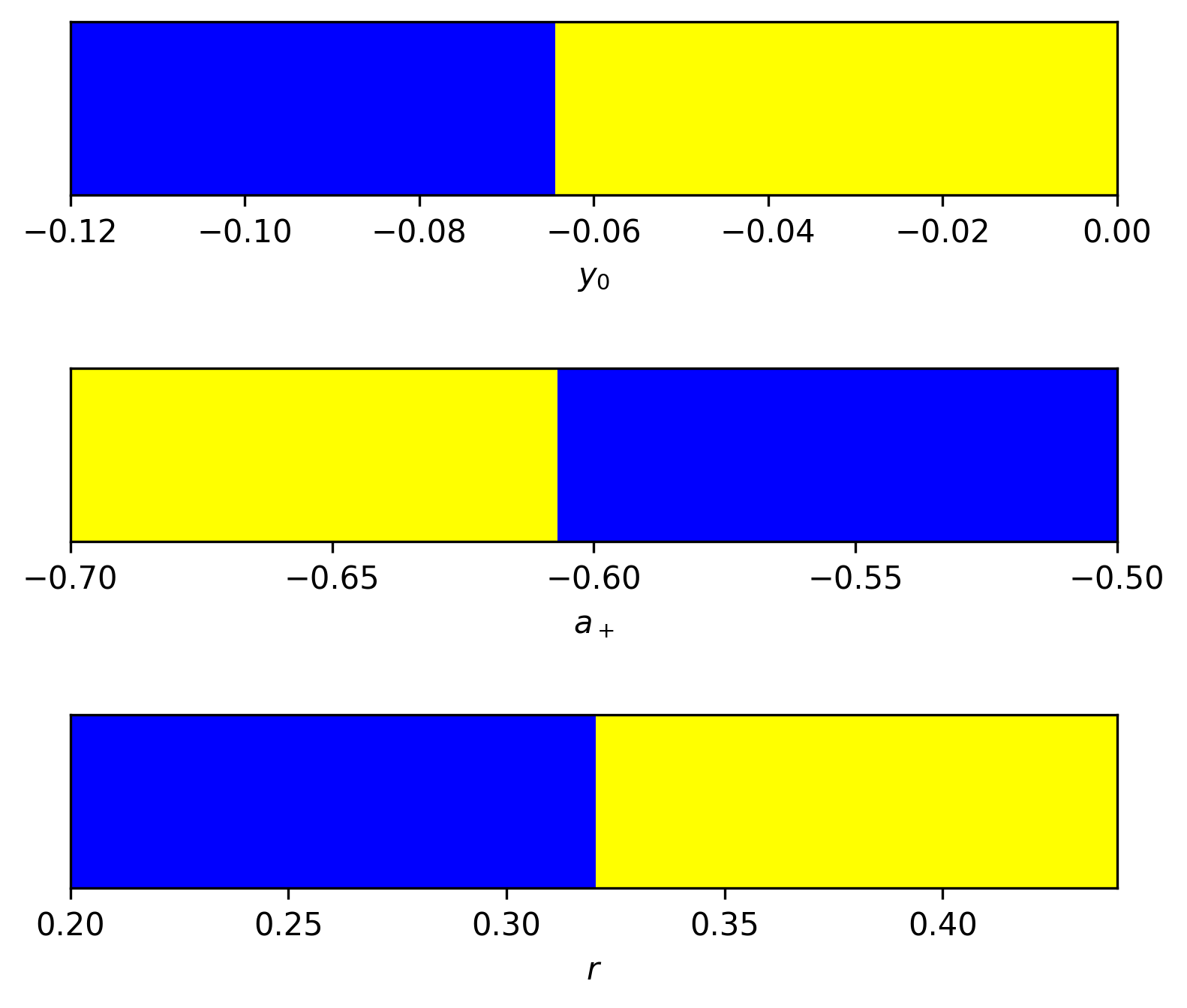}
    \end{subfigure}
    \caption{\justifying 
    Smooth basin boundary of the end fixed point yields smooth boundaries in the R-tipping region in the parameter space as well as in the rate space. (a) The basin of attraction (in blue) boundary of the fixed point corresponding to $(a_+, b_+) = (-0.68, -0.75)$. Initial conditions in yellow diverge to infinity, with $(x_n, y_n) \to (+\infty, -\infty)$ as $n \to +\infty$. (b) Plot of $\phi(r,x_0;\Lambda(s))$ in Eq.~(\ref{eq.phi}) -- with blue corresponding to $\phi=$ track and yellow to $\phi=$ tip (to infinity) --  as a function of different parameters (top to bottom): initial conditions $(x_0=0.5,y_0)$, corresponding to the basin-of-attraction of the fixed point for the end parameter $(a_+,b_+)=(-0.68, -0.75)$; the end parameter value $(a_+,b_+=-0.75)$, for a fixed starting parameter $(a_0,b_0)=(0.20, 0.25)$; rate $r$ using $(a_0,b_0)=(0.20, 0.25)$ and $(a_+,b_+)=(-0.68, -0.75)$ under parameter path Eq.~\eqref{eq:hennonDrift}.
    }
    \label{fig:HenonSmooth}
\end{figure}

\section{Computation of the fractal dimension}\label{app.fractal}

The fractal dimension of the different sets of interest are computed using the uncertainty algorithm described in Refs.~\cite{McDondald1985,lai_transient_2011} and implemented in our repository~\cite{wang_fractal_tipping_2026}. We begin by defining a one-dimensional line segment ($d=1$) in the space of interest: the phase-space of initial conditions $x_0$ (to compute the phase-space fractal), the parameter space $\lambda$ (to compute the parameter-space fractal), or the rate space $r$ (to compute the rate-space fractal). Here we consider this line parametrized by $y$, with $y\in[y_{a},y_{b}]$. We then compute the fractal co-dimension~$\alpha$ with the following algorithm:

\begin{enumerate}
\item For a given scale $\varepsilon < |y_b-y_a|$, repeat the following steps until a pre-defined number $M_u$ of uncertain points is obtained:
\begin{enumerate}
    \item Pick randomly with uniform probability a value $y\in[y_a+\varepsilon,y_b-\varepsilon]$ and  neighbouring points $y_\pm=y \pm \varepsilon$.
    \item Iterate the dynamical system~(\ref{eq.map}) for $y$ and $y_\pm$ and compute the value of the tipping function $\phi(y)$ and $\phi(y_\pm)$.
    \item If $\phi(y) \neq \phi(y_\pm)$, we consider $y$ to be uncertain ($M_u \mapsto M_u+1$).
\end{enumerate}
\item Compute the fraction of uncertain points $f(\varepsilon) = M_u/M,$ where $M$ is the number of times the previous loop was performed (i.e., the number of tested points).
\item Reduce $\varepsilon$ (e.g., $\varepsilon \mapsto \varepsilon/2$) and repeat the computation above.
\end{enumerate}
The uncertainty exponent $\alpha$ is estimated as the scaling exponent between $f(\varepsilon)$ and  $\varepsilon$. The fractal dimension $D$ of the underlying set (i.e., the boundary between $\phi=$ track and $\phi=$ tip intersected transversely by $y$), in the limit of small $\varepsilon$, satisfies
$$ D=1-\alpha.$$
We estimate $\alpha$ along a one-dimensional curve (a straight line) intersecting transversely the fractal set of interest, thus $0 \le \alpha \le 1$, and the uncertainty exponent $\alpha$ can thus be interpreted as the co-dimension of the fractal sets of interest.

The points shown in our numerical computations -- Figs.~\ref{fig:combined}, \ref{fig.henon3}, and \ref{fig:fractal_fits_pen} --  were estimated using $M_u=1,000$ in Fig.~\ref{fig:combined} and $M_u=150$ in Figs.~\ref{fig.henon3} and \ref{fig:fractal_fits_pen}. Error bars are computed assuming a negative binomial distribution. The slopes of the fitted lines correspond to the uncertainty exponent $\alpha$ and were estimated through linear regression. The associated $95\%$ confidence intervals were obtained by repeating the fit after bootstrapping the points.

\section{Tipping in the H\'enon map with smooth boundaries}\label{app.henonsmooth}

Figure~\ref{fig:HenonSmooth} shows an example of a smooth (non-fractal) boundary of the tipping function $\phi$ in cases in which the attractor $x^*$ at $\lambda_+$ has a smooth basin boundary. 


%

%

\end{document}